\documentclass[twocolumn,twocolappendix,numberedappendix,appendixfloat]{aastex701}

\usepackage{graphicx}
\usepackage{amsmath}
\usepackage{amssymb}
\usepackage{siunitx}
\usepackage{comment}
\usepackage{multirow}
\usepackage{xcolor}

\newcommand{\gtorder}{\mathrel{\raise.3ex\hbox{$>$}\mkern-14mu
             \lower0.6ex\hbox{$\sim$}}}
\newcommand{\ltorder}{\mathrel{\raise.3ex\hbox{$<$}\mkern-14mu
             \lower0.6ex\hbox{$\sim$}}}

\newcommand{\rh}{\hat{r}}
\newcommand{\Rh}{\hat{R}}
\DeclareMathOperator{\arcsech}{arcsech}
\DeclareMathOperator{\arcsectrig}{arcsec}

\begin{document}

\title{Dynamical Systematics for Time Delay Lenses and the Impact on the Hubble Constant}

\author[0000-0002-6482-2180]{R.~For\'es-Toribio}
\email{forestoribio.1@osu.edu}
\affiliation{Department of Astronomy, The Ohio State University, 140 West 18th Avenue, Columbus, OH 43210, USA}
\affiliation{Center for Cosmology and Astroparticle Physics, The Ohio State University, 191 W. Woodruff Avenue, Columbus, OH 43210, USA}

\author{C.~S.~Kochanek}
\email{}
\affiliation{Department of Astronomy, The Ohio State University, 140 West 18th Avenue, Columbus, OH 43210, USA}
\affiliation{Center for Cosmology and Astroparticle Physics, The Ohio State University, 191 W. Woodruff Avenue, Columbus, OH 43210, USA}

\author[0000-0001-9833-2959]{J. A. Mu\~noz}
\email{}
\affiliation{Departamento de Astronom\'{\i}a y Astrof\'{\i}sica, Universidad de Valencia, E-46100 Burjassot, Valencia, Spain}
\affiliation{Observatorio Astron\'omico, Universidad de Valencia, E-46980 Paterna, Valencia, Spain}

\correspondingauthor{Raquel For\'es-Toribio} \email{forestoribio.1@osu.edu}

\begin{abstract}
While time-delay lenses can independently probe $H_0$, the
estimates are degenerate with the convergence of the lens near
the Einstein radius. Velocity dispersions, $\sigma$, can be used
to break the degeneracy, with uncertainties
$\Delta H/H_0 \propto  \Delta\sigma^2/\sigma^2$ ultimately limited
by systematic uncertainties in the kinematic
measurements - measuring  $H_0$ to 2\% requires
$\Delta\sigma^2/\sigma^2 < 2\%$. Here we explore a broad range
of potential systematic uncertainties affecting eight time-delay lenses used in cosmological analyses. We find that:
(1) The characterization of the PSF in both absolute scale and
shape is important, with biases in $\Delta\sigma^2/\sigma^2$ up to $1$-$5\%$ for ground-based observations.
Small miscenterings of the lens are less important.
(2) The difference between the measured velocity dispersion
and the mean square velocity needed for the Jeans equations
is important, with up to $\Delta\sigma^2/\sigma^2 \sim 2$-$6\%$.
(3) The choice of anisotropy models is important with maximum changes of $\Delta\sigma^2/\sigma^2 \sim2$-$18\%$. Biases may be minimized by using models that reproduce the $h_4$ velocity moments typical of early-type galaxies.
(4) Small differences between the true stellar mass distribution
and the model light profile matter ($\Delta\sigma^2/\sigma^2 \sim 1$-$10\%$),
with radial color gradients further complicating the problem. The Jeans equations for mixed stellar populations imply that the correct profile is a population line equivalent width weighting corresponding to no broad band filter profile.
Finally, the homogeneity of the early-type galaxy
population means that many dynamically related parameters must be marginalized
over the lens sample as a whole and not over individual lenses.
\end{abstract}
\keywords{Gravitational lensing (670), Strong gravitational lensing (1643), Cosmological parameters (339), Extragalactic Distance Scale}

\section{Introduction}

Measurements of $H_0$ from time delays scale suffer from the degeneracy that 
$H_0 \propto 1 - \kappa_E$ (\citealt{Kochanek2002}, \citealt{Kochanek2006})
where a fundamental mathematical degeneracy means
that no differential lens data (positions, fluxes, etc.) can determine the local
convergence $\kappa_E$ near the Einstein radius, $R_E$ (see, e.g., \citealt{Gorenstein1988},
\citealt{Kochanek2002}, \citealt{Kochanek2006}, \citealt{Schneider2013},
\citealt{Wertz2018}, \citealt{Sonnenfeld2018}).  The properties of the radial mass distribution
that are determined by such data are the Einstein radius $R_E$ and the
dimensionless quantity $\xi = R_e \alpha''(R_E)/(1-\kappa_E)$ where 
$\alpha''(R_E)$ is the second derivative of the deflection profile at
$R_E$ (\citealt{Kochanek2020}).  The mathematical structure of the mass
model then determines $\kappa_E$ given the available constraints on 
$R_E$ and $\xi$ and the amount of freedom in the mass model.  

The two parameter or effectively two parameter mass models that are in
common use, lead to a unique value for $\kappa_E$ given $R_E$ and $\xi$.
For example, the power law model with $\alpha(r) = b^{n-1} r^{2-n}$
has $R_E=b$, $\xi=2(n-2)$ and $\kappa_E = (3-n)/2 = (2-\xi)/4$. 
While it is frequently said that lenses prefer density distributions
similar to the singular isothermal sphere with $n \simeq 2$  
(e.g., \citealt{Rusin2005}, \citealt{Gavazzi2007}, \citealt{Koopmans2009},
\citealt{Auger2010}, \citealt{Bolton2012}), the real constraint is
that $\xi \simeq 0$, which the power law model just happens to produce
for $n=2$.  This property makes it very dangerous to use lensing
data that strongly constrain $\xi$ in mass models with too few
degrees of freedom because they force the model to a particular
value of $\kappa_E$ and an estimate $H_0$ that is very precise but
inaccurate.  

In \cite{Kochanek2020,Kochanek2021}, we extensively demonstrate these
points and find that the accuracy of present estimates of $H_0$   
from lens time delays is at best $\sim 5$-$10\%$ regardless of the 
reported precision of the measurements.  The only way to avoid this
problem is to use mass models with more degrees of freedom so that
the relationship between $\xi$ and $\kappa_E$ is not one-to-one,
with the obvious consequence of larger uncertainties. For example, \cite{Birrer2020} found that $H_0$ is constrained to only $\sim$8\% for models with minimal assumptions about the mass profiles of seven time-delay lenses. Since \cite{Birrer2020} assumed a fully covariant formalism, the constraints were not averaged and, in consequence, the $H_0$ uncertainty in their work is larger than in \cite{TDCOSMO2025}. However, since the
fundamental problem is related to systematic uncertainties in the
structure of galaxies and their dark matter halos, averaging results
from multiple lenses will not necessarily lead to any improvements
in the accuracy of the results.

The alternative to constraining the mass models with lensing data is
to obtain  stellar dynamical measurements, usually the central velocity 
dispersion (e.g., \citealt{Grogin1996}, \citealt{Romanowsky1999}, \citealt{Treu2002}). In some cases there are 2D velocity dispersion measurements (\citealt{Shajib2026,Sheu2026}), but we will focus on central dispersion measurements for simplicity.
In purely stellar dynamical measurements of spherical systems, the distribution 
of the mass is determined by the changes in the observed stellar velocity
dispersions with radius.  However, as illustrated by \cite{Binney1982} and
\cite{Tonry1983}, the radial distribution of mass and the radial variation
of the orbital anisotropy are strongly degenerate.  The degeneracy can be
broken if the complete line-of-sight (los) velocity distribution is known
(e.g., \citealt{Dejonghe1987}, \citealt{Merritt1987},  \citealt{Dejonghe1992},
\citealt{Merritt1993}, \citealt{Gerhard1993}).  While this is impractical
for extragalactic studies, measuring higher order moments of the 
velocity distribution supplies most of the necessary information. In
most cases, these moments are characterized by a Gauss-Hermite 
polynomial expansion (\citealt{vanderMarel1993}), where typical
amplitudes for the fourth moment in elliptical galaxies 
are $h_4 \simeq 0.01$ to $0.04$ (e.g., \citealt{Emsellem2004},
\citealt{Arnold2014},
\citealt{Veale2017}). Of course, as suggested by
the designation of elliptical, early-type galaxies are not spheres and they can also
have non-zero odd velocity moments (bulk rotation and $h_3$),
but we defer this issue until the discussion.  
 
Combining stellar dynamics with lensing avoids some of these stellar
dynamical problems because the lens geometry very accurately constrains 
the mass enclosed by the Einstein ring.  
However, as emphasized by \cite{Kochanek2006}, it introduces a problem which is
of little concern in stellar dynamical studies -- in lensing, the inference about
the mass distribution depends on the {\it absolute} calibration of the velocity
dispersions because it comes from comparing the stellar dynamical and lensing
masses.   In stellar dynamical studies, inferences about the mass distribution
depend only on ratios of velocity dispersions.  As a result, small calibration
or other systematic
errors that would be relatively unimportant for a purely stellar dynamical analysis
can be very important in a lensing analysis.  This is particularly true as the
fractional uncertainties become smaller.    

\begin{figure}
\centering
\includegraphics[width=0.47\textwidth]{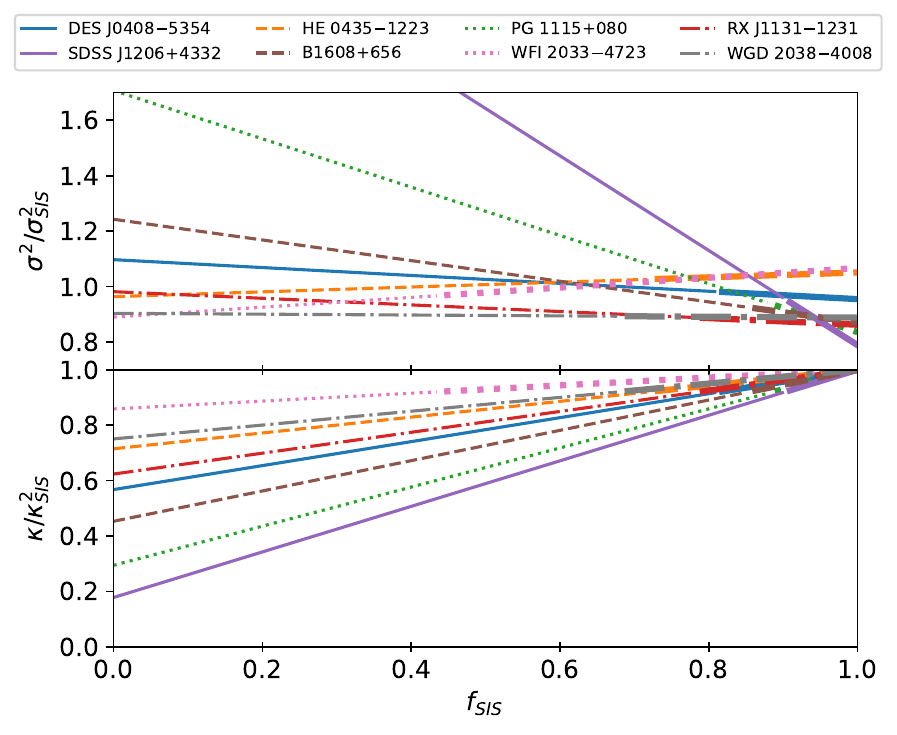}
\caption{
  Observed central velocity dispersion $\sigma^2/\sigma_{SIS}^2$ (top) and surface density
  at the Einstein radius $\kappa_E/\kappa_{SIS}$ normalized by the values for an 
  SIS lens model.  The heavier lines indicate the changes that will increase $H_0$
  by 8\% from the SIS model.  
  The curves are for the eight time-delay lenses with the parameter values from Table~\ref{tab:data}. The velocity dispersion includes a Gaussian
  PSF model and the extraction aperture.   
  The estimated Hubble constant increases as $-\kappa/2\kappa_{SIS}$ as the 
  surface density decreases.
  }
\label{fig:model}
\end{figure}

Figure~\ref{fig:model} illustrates this problem with a simple model (see Appendix \ref{app:model}) consisting
of a \cite{Hernquist1990} model for the stars combined with a one parameter $0~\leq 
~f_{SIS}~\leq~1$
mass distribution that smoothly shifts from a constant mass-to-light ($M/L$) ratio model for
$f_{SIS}=0$ to a singular isothermal sphere (SIS) with a flat rotation curve at
$f_{SIS}=1$.  Models of gravitational lenses generally find (but recall the earlier discussion) that the mass distribution
is similar to the SIS model (e.g., \citealt{Munoz2001}, \citealt{Rusin2005}, \citealt{Gavazzi2007}, \citealt{Koopmans2009},
\citealt{Auger2010}), possibly with some evolution with redshift (e.g., \citealt{Bolton2012})
and correlations with mass (e.g., \citealt{Auger2010}).
These are the observed velocity dispersions for the eight lenses in
Table~\ref{tab:data} that were used in cosmological studies, assuming the listed scale lengths 
$s$ for the Hernquist model and a Gaussian model for the point spread function (PSF).
Figure~\ref{fig:model} shows both the
velocity dispersion and the surface density at the Einstein ring relative to an SIS 
model as a function of $f_{SIS}$.  As the dark matter fraction increases, the mass
is less centrally concentrated so the velocity dispersion drops and the surface density
at the Einstein ring increases, leading to a decrease in $H_0$.
 
Since the curves in Figure~\ref{fig:model} are straight lines, we can characterize
each lens by its sensitivity to changes in the velocity dispersion. The convergence (surface
density) of the SIS model at the Einstein radius is $\kappa_{SIS}=1/2$ and 
$H_0 \propto 1-\kappa$, so we define
\begin{equation}
     \psi = \frac{ d H_0/H_{SIS}}{d \sigma_*^2/\sigma_{SIS}^2} 
       = -\frac{1}{2} \frac{d \kappa/\kappa_{SIS}}{ df_{SIS} }
         \frac{ d f_{SIS}}{d \sigma_*^2/\sigma_{SIS}^2 }.
\end{equation}to characterize the fractional sensitivity of $H_0$ estimates to 
fractional changes in the dispersion.
The slopes are $\psi=$ 1.52, $-$1.63, 0.40, 1.58, 0.24, 0.73, $-$0.40, and 8.15 for the eight time-delay lenses we considered: DES~J0408$-$5354, HE~0435$-$1223, PG~1115$+$080, RX~J1131$-$1231, SDSS~J1206$+$4332, B1608$+$656, WFI~2033$-$4723 and WGD~2038$-$4008, respectively. The median absolute slope is 1.13 meaning that a 1\% change in $\sigma_*^2$ roughly leads to a 1\% change in $H_0$.
We are not too concerned about the particular numbers -- they simply provide a 
rough indication of the sensitivity of the cosmological results to any 
systematic shifts in the dynamics.  Since an uncertainty below 2\% is needed for $H_0$ measurements from lenses to be competitive, the systematic uncertainties in $\sigma_*^2$ ($\sigma_*$) can be no higher than 2\% (1\%).

In practice, many time-delay lens models show far less sensitivity to changes in the
dynamical constraints.  Some specific examples
are that \cite{Suyu2010} and \cite{Wong2017} find that switching between 
\cite{Hernquist1990} and \cite{Jaffe1983} models for the distribution of the stars
changes $H_0$ by less than 1\%.  In \cite{Millon2020}, they estimate that
$\psi \simeq 0.06$ instead of unity. This insensitivity is an indication that the dynamical constraints
are playing no role in the results.  The lens data so tightly constrains
the mass models that changes in the velocity dispersion have negligible effects
on the results, as illustrated in one of the example problems shown in 
\cite{Kochanek2020}. Treating the mass-sheet degeneracy explicitly should increase the response of the $H_0$ estimates to the kinematic constraints \citep[see, e.g.][]{Birrer2020,Shajib2023,TDCOSMO2025}.

Recently, there have been efforts on quantifying and minimizing the systematics produced by the dynamical constraints. \cite{Knabel2025} made a detailed study of the uncertainties due to the choices of stellar templates used in determining the velocity dispersion. \cite{Huang2026} studied the impact of projection and selection effects of the lens galaxies on the inferred Hubble constant. \cite{Liang2025ML} study the biases produced when assuming a constant stellar mass to light ratio. \cite{Liang2025ani} explore some of the systematic biases produced by the anisotropy model assumption. And \cite{Verma2026} quantify the systematic biases arising from the choice of the anisotropy profile when incorporating 2D kinematic information.

In this paper we consider a broader range of systematic problems in using dynamical constraints when estimating
$H_0$ from gravitational lenses. 
In \S\ref{sec:measure} we consider two issues associated with measuring
the velocity dispersion: (1) the model for the point spread function (PSF), and (2)
the difference between the velocity dispersion appearing in the Jeans equation
and the velocity dispersion measured from the spectrum. In \S\ref{sec:isotropy}
we consider the biases arising from the choice of  anisotropy profile on the velocity dispersion.
In \S\ref{sec:profile} we examine the consequences of small mismatches between the adopted stellar mass profile and the actual density distribution. In \S\ref{sec:homogeneous} we discuss how the treatment of the lens galaxies as
a completely heterogeneous population will significantly underestimate the
final uncertainties to the extent that early-type galaxies are a 
homogeneous population in both dynamics and stellar populations.  We discuss our arguments and provide future guidance in
\S\ref{sec:disc}.

\section{Measuring the Velocity Dispersion}
\label{sec:measure}

We assume a spherically symmetric lens galaxy and that
the template star spectra used to analyze the spectrum are noiseless and have perfectly
understood spectral resolutions. Breaking any of these assumptions will simply
add additional sources of systematic uncertainty.

The standard dynamical analysis models the spectrum by convolving a
template star spectrum with a Gaussian and fitting it to the data combined
with a polynomial or equivalent to remove any residual large scale spectral
differences.
The best fit dispersion is then corrected for the line spread function (LSF)
by subtracting (usually) the width of the night sky lines in quadrature from the
raw measurement.  The resulting dispersion measurement $\sigma_*$ is then
modeled using the Jeans equations to constrain the 
gravitational lens model. This dynamical model requires a model for
the mass distribution, the stellar distribution and the orbital anisotropy.  
It must also include a model for the seeing (point spread function, PSF) at the time
of the observations and the extraction aperture in order to obtain the correct
average over the lens galaxy. In the following subsections we study the possible systematics that can arise in modeling the point spread function (Sect.~\ref{sec:PSF}) and in the assumptions made when applying the Jeans equation (Sect.~\ref{sec:disp}). The possible systematic errors in the LSF model were considered but they are negligible compared to other sources of systematics.

\begin{table*}
  \centering
  \caption{Summary of dispersion observations for the time-delay lenses.}
  \begin{tabular}{lDr@{$\pm$}lcr@{$\times$}lccccc}
  \hline
  \hline
  \multicolumn{1}{c}{Lens}    &
  \multicolumn{2}{c}{$z_{\text{lens}}$}   &
  \multicolumn{2}{c}{$\sigma_{\ast}$}   &
  \multicolumn{1}{c}{Seeing} &
  \multicolumn{2}{c}{Aperture}   &
  \multicolumn{1}{c}{Instrument} &
  \multicolumn{1}{c}{$R_e$} &
  \multicolumn{1}{c}{$q$}   &
  \multicolumn{1}{c}{$s$}  &
  \multicolumn{1}{c}{$\theta_E$} \\
  \multicolumn{1}{c}{}    &
  \multicolumn{2}{c}{}    &
  \multicolumn{2}{c}{(km/s)}   &
  \multicolumn{1}{c}{FWHM (\arcsec)} &
  \multicolumn{2}{c}{(\arcsec$^2$)}   &
  \multicolumn{1}{c}{} &
  \multicolumn{1}{c}{(\arcsec)}   &
  \multicolumn{1}{c}{}   &
  \multicolumn{1}{c}{(\arcsec)}   &
  \multicolumn{1}{c}{(\arcsec)}   \\
\hline
\decimals

DES~J0408$-$5354  &0.597 &242.3&12.2 &0.52 &$\pi$&0.5$^2$   &MUSE    &1.940 & 0.80 &1.067 &1.92 \\
HE~0435$-$1223    &0.455 &226.6&5.8  &0.15 &0.55&0.55       &NIRSpec &1.800 & 0.93 &0.990 &1.22 \\
PG~1115$+$080     &0.311 &235.7&6.6  &0.15 &0.55&0.55       &NIRSpec &0.450 & 0.95 &0.248 &1.08 \\
RX~J1131$-$1231   &0.295 &303.0&8.3  &0.15/0.96 &$\pi$&0.955$^2$ &NIRSpec/KCWI &1.910 & 0.94 &1.050 &1.63 \\
SDSS~J1206$+$4332 &0.745 &290.5&9.5  &0.15 &0.55&0.55       &NIRSpec &0.290 & 0.85 &0.160 &1.25 \\
B1608$+$656       &0.630 &305.3&11.0 &0.15 &0.55&0.55       &NIRSpec &0.590 & 1.00 &0.324 &0.81 \\
WFI~2033$-$4723   &0.657 &210.7&10.5 &0.15 &0.55&0.55       &NIRSpec &1.970 & 0.83 &1.084 &0.94 \\
WGD~2038$-$4008   &0.228 &254.7&16.3 &1.26 &$\pi$&0.75$^2$  &MUSE    &2.223 & 0.85 &1.223 &1.38 \\
\hline
\end{tabular}
\tablecomments{We list the measured velocity dispersion, $\sigma_*$, the seeing reported as the full width at half maximum (FWHM), the extraction aperture, the instrument used to acquire the spectra, the intermediate axis half-light radius of the light profile, $R_e$, the axis ratio of the light profile, $q$, the Hernquist intermediate scale radius, $s=0.55R_e$, and the Einstein radius, $\theta_E$.}
\label{tab:data}
\end{table*}

\subsection{Observations}\label{sec:obs}

Table~\ref{tab:data} summarizes the most up-to-date dynamical observations for eight lenses used to jointly infer $H_0$ in \cite{TDCOSMO2025}. These  measurements were acquired with integral field units (IFUs) and the measured velocity dispersions are spatially integrated within the apertures with the exception of RX~J1131$-$1231 which has radially resolved kinematics. To homogenize the analysis, we consider for this lens the recovered velocity dispersion within an aperture of half of the effective radius and for the NIRSpec and KCWI observations. Six of the lenses were previously modeled by H0LiCOW (\citealt{Wong2020}) with ground-based long-slit spectra. DES~J0408$-$5354 was modeled by \cite{Shajib2020} and later included in the TDCOSMO hierarchical $H_0$ joint inference with the other previous six lenses (\citealt{Birrer2020}). Lastly, WGD~2038$-$4008 was modeled by \cite{Shajib2023} and \cite{Wong2024} inferred $H_0$ from this single system. A summary of the results obtained with the previous ground-based kinematic measurements of these lenses can be found in Appendix \ref{app:prev}.

Several different photometric
models are used to model the light profiles of the lens galaxies, but generally the lens galaxies are fitted using one or more elliptical \cite{Sersic1968} profiles
\begin{equation}
      \Sigma (R) = A \exp\left( -b(n) \left[\left(R/R_e\right)^{1/n}-1\right] \right),
      \label{eqn:sersic}
\end{equation} elliptical \cite{Hernquist1990} profiles
\begin{equation}
\Sigma (R) = \frac{A}{\left(s^2-R^2\right)^2}\left[\left(2s^2+R^2\right)X(R) - 3 \right]
\end{equation} where
\begin{equation}
X(R)  =
\begin{cases} 
     \frac{1}{\sqrt{1-(R/a)^2}}\arcsech (R/s) &R\le s \\
     \frac{1}{\sqrt{(R/a)^2-1}}\arcsectrig (R/s) &R\ge s
   \end{cases}
\end{equation} or elliptical pseudo-Jaffe profiles
\begin{equation}
\Sigma (R) = A \left[\left(R^2 + s^2 \right)^{-1/2} - \left(R^2 + a^2 \right)^{-1/2} \right].
\end{equation}
For all profiles, an elliptical coordinate transformation can be applied as $R^2 \rightarrow x^2 + y^2/q^2$  with an axis
ratio of $q\le1$ and a major axis effective radius of $R_e$.  
The \cite{Sersic1968} profiles include the exponential
disk profile as $n=1$ and the \cite{deVaucouleurs1948} profile as $n=4$.  The
dimensionless quantity $b(n)$ is defined so that half the light lies inside the
elliptical isophote with a major axis radius of $R_e$. The effective radius of the \cite{Hernquist1990} profile in units of the scale radius is $R_e \simeq 1.82 s$. There is no simple
expression for the effective radius of the pseudo-Jaffe profile, but
in the limit of no core radius ($s\rightarrow 0$) it is $R_e=3a/4$.

The light profiles used for the time-delay lenses in \cite{TDCOSMO2025} are new fits with a single S\'ersic profile for DES~J0408$-$5354, double S\'ersic profiles for HE~0435$-$1223, PG~1115$+$080, RX~J1131$-$1231, SDSS~J1206$+$4332, and WFI~2033$-$4723, a triple S\'ersic profile for WGD~2038$-$4008 and a Hernquist profile is used for B1608$+$656.\footnote{The light profile parameters for the lenses can be found in \url{https://github.com/TDCOSMO/TDCOSMO2025_public/blob/main/TDCOSMO_sample/tdcosmo_sample.yaml}} \cite{TDCOSMO2025} report the intermediate axis effective radii (the geometric mean of the major and minor axes) as $R_e$. In our following analyses we will use this intermediate axis radius.

\cite{TDCOSMO2025} use the photometric fit to build the dynamical model. However, most of the previous models did not use the actual photometric model for their
stellar dynamical calculations.  Instead, they assume that the stellar
density can be modeled by a spherical \cite{Hernquist1990} distribution 
\begin{equation}
    \rho(r) = \frac{1}{2 \pi} \frac{ M s}{r (r+s)^3 }.
       \label{eqn:hernrho}
\end{equation}
They also sometimes use a \cite{Jaffe1983} distribution
\begin{equation}
    \rho(r) = \frac{ 1}{ 4 \pi } \frac{ M a}{ r^2 (r+a)^2 }.
\end{equation} In three dimensions, the pseudo-Jaffe
density distribution of $\rho \propto (r^2+s^2)^{-1}(r^2+a^2)^{-1}$
is very similar to the \cite{Jaffe1983} model.
Where the photometric
model uses multiple \cite{Sersic1968} or pseudo-Jaffe components, the
effective radius of the combined distribution is used to set the scale
for the \cite{Hernquist1990} model.
Since we are simply illustrating sources of systematic errors, we use Hernquist models for the light profiles with a scale length $s=0.55R_e$ set to the observed estimate from \cite{TDCOSMO2025} for simplicity (see Table~\ref{tab:data}).

Depending on the calculation, we use different levels of approximation
for the combined effects of the extraction aperture and the seeing. In the following subsections (\ref{sec:PSF} and \ref{sec:disp}), we both consider the full convolution with
the PSF and use
the actual extraction geometry.  In \S\ref{sec:isotropy} and \S\ref{sec:profile} we approximate the
effects of seeing by broadening the extraction aperture by the
dispersion $\sigma = FWHM/(8\ln 2)^{1/2}$ of a Gaussian model for the 
seeing, and use a circular aperture with this area.  
For square apertures with a $\Delta x \times \Delta y$ extraction
region, this becomes a radius of 
$\left[ (\Delta x + 2\sigma)(\Delta y + 2\sigma)/\pi\right]^{1/2}$.
Since we are generally making differential comparisons, this approximation has little effect on the results.

\subsection{The Point Spread Function}\label{sec:PSF}

Most of the velocity dispersion measurements treat the point spread
function as a Gaussian with a given FWHM. Three issues to consider are (1) the
necessary accuracy with which the width of the PSF must  
be known, (2) the precision with which the aperture must be 
centered on the galaxy, and (3) whether a Gaussian
is an adequate model for the PSF.

For these calculations we adopt a dynamical model close to the one used in \cite{TDCOSMO2025}. We assume an SIS mass model, a Hernquist stellar distribution with a scale radius $s$ given in Table~\ref{tab:data} and a constant orbital anisotropy, $\beta$, ranging from $-$0.25 (slightly tangential) to 0.5 (moderately radial) which covers the isotropic case, $\beta=0$. The range is selected to roughly match \cite{TDCOSMO2025} ($\beta\in[-0.2544,0.2431]$) while also exploring the ranges found in other studies \citep[see, e.g.,][]{Gerhard2001,Koopmans2009}. Once the los velocity dispersion, $\sigma_{\text{los}}$, is obtained for each model, the observed velocity dispersion for a centered aperture is computed for each system as
\begin{equation}\label{eq:sigobs}
    \sigma^2_{\text{obs}}=\frac{\int_0^\infty \Sigma(R)\,\sigma^2_{\text{los}}(R)\,\text{AP}(R)\,R\,dR}{\int_0^\infty \Sigma(R)\,\text{AP}(R)\,R\,dR},
\end{equation}where $\Sigma(R)$ is the stellar surface mass density and $\text{AP}(R)$ is the angular average of the convolution of the aperture with the PSF using the parameters and apertures listed in Table~\ref{tab:data}.

The sources of systematic errors studied here are more significant for the ground-based measurements (namely, the measurements with MUSE and KCWI) owing to the larger seeing. While PSF modeling issues are likely smaller for spacecraft observations, the kinematic results may also be more sensitive to the details of modeling the central regions, as we encounter with some unexpected trends discussed below. \cite{TDCOSMO2025} model the PSFs of NIRSpec and KCWI as Gaussians with the FWHM given in 
Table~\ref{tab:data}. The PSFs of DES~J0408$-$5354 and WGD~2038$-$4008 were characterized as Moffat profiles but the exponent parameter is not reported so we simply assume Gaussian profiles.

\begin{figure}
\centering
\includegraphics[width=0.47\textwidth]{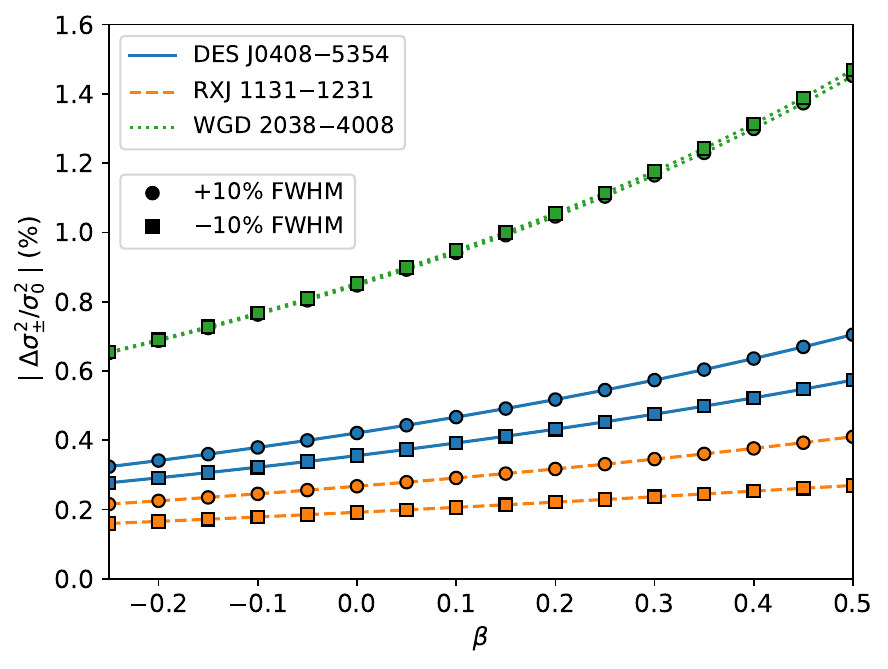}
\caption{Absolute values of the fractional changes, $\left|\Delta\sigma^2_{\pm}/\sigma^2_0\right|$, due to misestimates of the seeing FWHM as a function of anisotropy, $\beta$. The curves with circles (squares) are the fractional changes if the 
  reported FWHM is increased (decreased) by 10\% assuming a Gaussian PSF. Each system  is plotted in a different color and line style. 
  }
\label{fig:fwhm}
\end{figure}

We first modeled the PSF
with either the reported seeing FWHM from Table~\ref{tab:data}
or a FWHM 10\% larger and smaller. For a FWHM larger (smaller) than the reported
value, the fractional change in the mean square velocity dispersion, $\Delta\sigma^2_{\pm}/\sigma^2_0 \equiv \left(\sigma^2_{\pm 10\% FWHM}-\sigma^2_{FWHM}\right)/\sigma^2_{FWHM}$, 
is negative (positive), as one would expect. In Figure~\ref{fig:fwhm} we show the absolute values to 
ease the comparison for both cases. We find the absolute largest 
fractional changes for
WGD~2038$-$4008 and the smallest for RX~J1131$-$1231.
The average changes are smallest for a slightly tangential ($\beta=-0.2$)
velocity distribution, with a range between the lens systems from 0.17 to 0.69\%, and then increases as the distribution
becomes isotropic and then radially anisotropic, with ranges of [0.18, 0.77]\%, [0.19, 0.85]\%, [0.21, 0.95]\%, [0.22, 1.06]\%, [0.24, 1.18]\%, [0.25, 1.31]\%, and [0.27, 1.47]\% for a 
constant anisotropy of $\beta=-0.1$, $0$, $0.1$, $0.2$, $0.3$, $0.4$, and 
$0.5$, respectively. If misestimates of the FWHM are random, this source of error can be reduced by averages over the lens sample. However, the error can also be systematically biased if introduced by the PSF measurement or modeling procedure.

\begin{figure}
\centering
\includegraphics[width=0.47\textwidth]{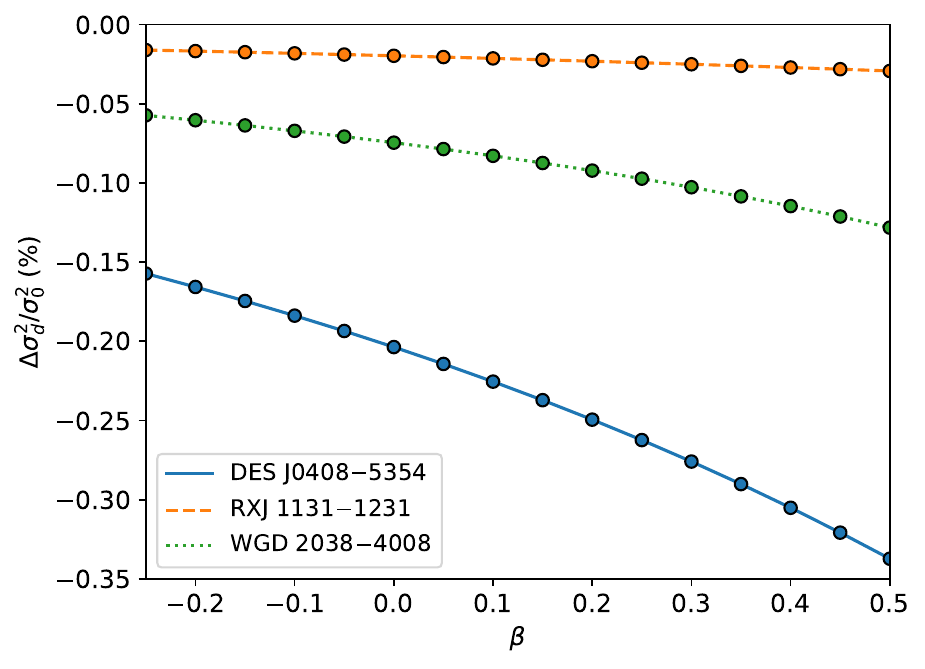}
\caption{
  Fractional changes, $\Delta\sigma^2_d/\sigma^2_d$, for apertures miscentered by half a pixel from the center of the galaxy as a function of anisotropy, $\beta$. Each system is plotted in a different color and line style, and we assumed a Gaussian PSF for all the cases.
  }
\label{fig:miscent}
\end{figure}

We also studied the effect of miscentering the aperture by half a pixel (0\farcs1, 0\farcs075, and 0\farcs05 for MUSE, KCWI, and NIRSpec, respectively). Since the velocity dispersion decreases with radius, miscentering the aperture leads to a lower measured value than if it 
is placed at the center. Hence, the fraction 
$\Delta\sigma^2_d/\sigma^2_d\equiv(\sigma^2_d-\sigma^2_0)/\sigma^2_0$ is 
negative, as seen in Figure~\ref{fig:miscent}.
In this case, DES~J0408$-$5354 has the largest differences and 
RX~J1131$-$1231 the smallest. The tangential $\beta=-0.2$ model has 
the smallest changes with a range from $-0.017$ to $-0.17$\%. As we 
increase the anisotropy, the differences become 
greater with ranges of [$-0.018$, $-0.18$]\%, [$-0.020$, $-0.20$]\%, [$-0.021$, $-0.23$]\%, [$-0.023$, $-0.25$]\%, [$-0.025$, $-0.28$]\%, [$-0.027$, $-0.31$]\%, and [$-0.029$, $-0.34$]\% for 
$\beta=-0.1$, $0$, $0.1$, $0.2$, $0.3$, $0.4$, and 
$0.5$, respectively.
Thus, a small miscentering has less impact 
on the velocity dispersion than small misestimates of the 
FWHM.

In addition, PSFs are not Gaussians (as adopted for NIRSpec and KCWI observations), instead they possess extended 
wings that can be traced out to several degrees from
a bright star (\citealt{King1971}).  Better analytic
models of the PSF are usually based on the \cite{Moffat1969}
profile
\begin{equation}
  I(R) \propto \left[ 1 + (R/R_0)^2 \right]^{-\eta},
\end{equation}where $R_0$ and $\eta$ are related to the FWHM by $\text{FWHM}=2R_0\sqrt{2^{1/\eta}-1}$.
\cite{Racine1996} found that a $\eta=4$ model worked
well over 7~mag from the peak. A Moffat profile was used in the analysis of the MUSE observations but the value of $\eta$ was not reported. The sum of two Moffat
profiles having the same FWHM with 80\% of the light 
in a $\eta=7$ profile and 20\% in a $\eta=2$ profile 
was a good fit over more than 15~mag from the peak.

\begin{figure}
\centering
\includegraphics[width=0.47\textwidth]{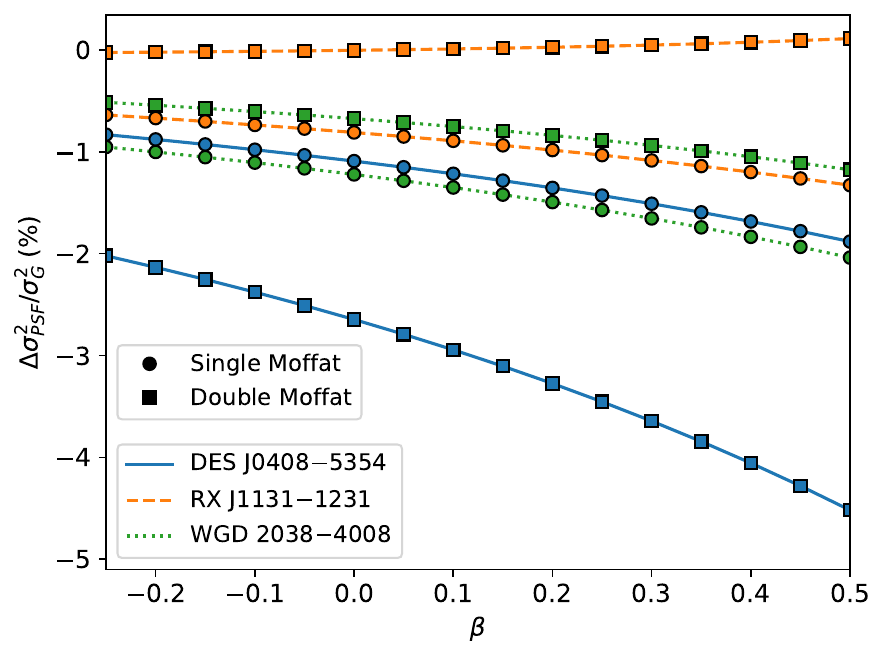}
\caption{
  Fractional changes of squared velocity dispersion relative to the value for a Gaussian PSF with the same FWHM for single (circles) and double (squares) Moffat profiles as a function of anisotropy, $\beta$. Each system is plotted in a different color and linestyle.
  }
\label{fig:psf}
\end{figure}

Figure~\ref{fig:psf} shows the
fractional differences in the square of the velocity
dispersion between the more realistic PSFs and 
Gaussians with the apertures reported in \cite{TDCOSMO2025} and as a
function of the constant anisotropy $\beta$.  
The more realistic PSFs include
more light from further out in the galaxy because of
their extended wings, leading to a lower measured
velocity dispersion than would be predicted by a 
Gaussian. Hence, using a Gaussian PSF to model 
data with a PSF more like these
models will also drive the mass distribution to be less
centrally concentrated than it should be. 

However, in the case of the double Moffat for RX~J1131$-$1231 the fractional changes are positive because the change in the numerator of Eqn.~\ref{eq:sigobs} is smaller than that in the denominator, even though both factors decrease with respect to the Gaussian PSF case as expected. The small change in the numerator is a consequence of the large steepness in velocity dispersions produced by the nonphysical behavior at the center from using an SIS potential. This can be indicating that a more sophisticated mass model is necessary to model precise velocity dispersion measurements, especially for the NIRSpec measurements were these inverted trends are more often encountered.

For a $\beta=-0.2$ velocity distribution, the fractional differences, $|\Delta\sigma^2_{PSF}|/\sigma^2_G$, are smaller, with a range for a single (double) Moffat 
of [$-0.67$, $-1.00$]\%
([$-0.54$, $-2.13$]\%). As we increase the anisotropy parameter, the 
fractional changes increase. For $\beta=-0.1$, $0$, $0.1$, $0.2$, $0.3$, $0.4$, and 
$0.5$, they rise to 
[$-0.74$, $-1.11$]\% ([$-0.60$, $-2.38$]\%),
[$-0.81$, $-1.22$]\% ([$-0.67$, $-2.65$]\%), 
[$-0.89$, $-1.35$]\% ([$-0.75$, $-2.94$]\%),
[$-0.98$, $-1.49$]\% ([$-0.84$, $-3.27$]\%),
[$-1.09$, $-1.65$]\% ([$-0.94$, $-3.64$]\%),
[$-1.20$, $-1.83$]\% ([$-1.05$, $-4.06$]\%), and
[$-1.33$, $-2.04$]\% ([$-1.18$, $-4.52$]\%), respectively.
We also find the largest differences for WGD~2038$-$4008 (DES~J0408$-$5354) when changing the PSF to a single (double) Moffat profile while RX~J1131$-$1231 has the smallest differences.

In summary, systematic uncertainties due to estimates and models of the PSF are likely present in ground-based observations with amplitudes comparable to the required $H_0$ precision for cosmological studies. Underestimates of the PSF width, miscentering the lens galaxy and assuming a Gaussian PSF lead to underestimates of the velocity dispersion which will drive models to have more extended dark matter and so lower values of $H_0$, while overestimating the PSF width has the reverse effect. This is likely less of a problem for the measurement of central velocity dispersion with space-based observations \citep[see, e.g.][for the JWST/NIRSpec PSF treatment]{Shajib2026} given that the fractional PSF errors compared to the scale length of the galaxy are smaller. However, at large distances from the center, the absolute PSF errors become more important and the measurements at large radii can be subjected to potential biases in 2D kinematic measurements.

\subsection{The Measured Dispersion}
\label{sec:disp}

Ignoring the additional complications of the effects of the finite aperture size and seeing, the measured velocity dispersion $\sigma_\ast$ is not the quantity $\sigma_r$ appearing
in the Jeans equations. The measured dispersion is the width of
the best fit Gaussian convolution kernel (corrected for the LSF, etc.) 
that, when combined with the LSF, broadens the features in the spectrum of a template 
star to match the data.  The quantity really needed for the Jeans equations is the root mean square velocity of the stars, $v_\text{rms}$.  These
two quantities are the same only if the velocity
distribution of the stars is a Gaussian.  For nearby galaxies,
the convolution kernel can be expanded as a Gauss-Hermite
polynomial (\citealt{vanderMarel1993}), where the next order
term beyond a simple Gaussian for a spherical non-rotating system is the fourth order term $h_4$. 
Radial (tangential) anisotropies  lead to $h_4 > 0$ ($h_4 < 0$).  \cite{vanderMarel1993} find a rough relation of $v_\text{rms} \simeq \sigma_* (1 +\sqrt{6}h_4)$, so the measured dispersion $\sigma_*^2$ is an underestimate (overestimate) of $v_\text{rms}$ if $h_4 > 0$ ($h_4 < 0$).

We investigate this for the Osipkov-Merritt (O-M, \citealt{Osipkov1979}, \citealt{Merritt1985})
model.  The O-M model has a radial anisotropy
profile of 
\begin{equation}
    \beta = \frac{r^2}{r^2+r_a^2}
    \label{eqn:OM}
\end{equation}where the orbits are 
isotropic near the center and radial in the halo with a 
break radius $r_a$ between the two regimes. Previous time-delay cosmography models typically marginalized the results for each lens over the range
$[0.5,5]R_e$, corresponding to $[0.9,9]s$.  
We use the O-M model instead of a constant anisotropy model (as in \citealt{TDCOSMO2025}) because in the O-M model
the distribution function (DF), $f(Q)$, depends only on the quantity
$Q = \epsilon - L^2/2 r_a^2$
where $\epsilon$ is the binding energy and $L$ is the angular 
momentum, instead of the two parameters separately, $f(\epsilon,L)$, as in other anisotropy models. This is for computational convenience -- the underlying issue is present for any anisotropy model. The form of $f(Q)$ is analytic for the \cite{Hernquist1990}
mass distribution and can be calculated
numerically if the stellar mass follows a Hernquist distribution 
but the underlying potential is an SIS distribution. To avoid 
divergences due to the SIS potential, we add an outer scale length $a$ 
so that the mass profile falls as $1/r^4$ for $r/a\gg1$ instead of as 
$1/r^2$. We set the outer scale to $a/s=1000$ so this has a negligible effect on the results.

\begin{figure}
\centering
\includegraphics[width=0.47\textwidth]{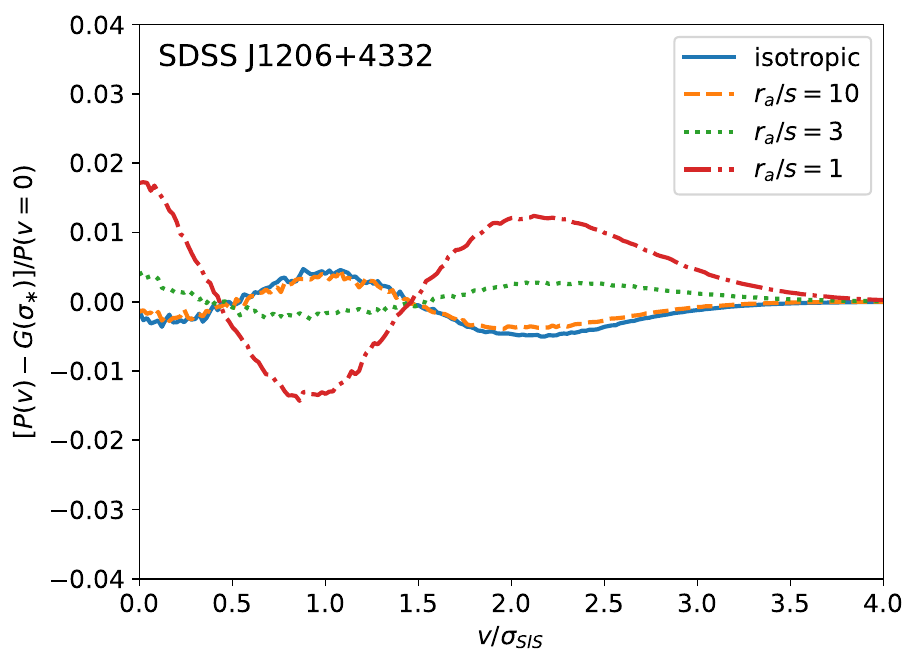}
\caption{
  The differences $[P(v)-G(\sigma_{\ast})]/P(v=0)$ 
  between the los velocity distribution ($P(v)$) and the best 
  fit Gaussian model ($G(\sigma_{\ast})$) normalized by the peak of
  the los velocity distribution at zero velocity, $P(v=0)$, 
  for SDSS~J1206$+$4332. The anisotropy radii considered are 
  $r_a/s\rightarrow \infty$ (isotropic), $10$, $3$, and $1$. 
  The velocity distribution accounts for Gaussian PSF effects and the rectangular extraction aperture from Table~\ref{tab:data} is applied. There is some noise 
  due to the Monte Carlo integration and sampling.
  }
\label{fig:DF-1206}
\end{figure}

\begin{figure}
\centering
\includegraphics[width=0.47\textwidth]{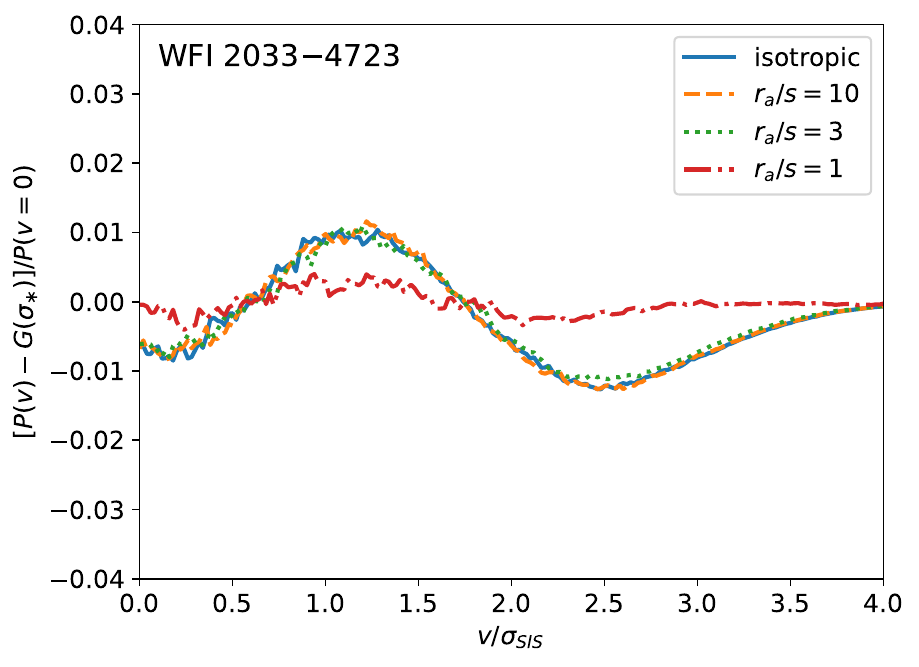}
\caption{
  The same as in Figure~\ref{fig:DF-1206} but for WFI~2033$-$4723.
  }
\label{fig:DF-2033}
\end{figure}

Figures~\ref{fig:DF-1206} and \ref{fig:DF-2033}  illustrate the differences
between the true los velocity distribution, $P(v)$, and the best fit Gaussian 
model, $G(\sigma_{\ast})$, for SDSS~J1206$+$4332 and WFI~2033$-$4723, respectively. The figures show the 
difference normalized by the peak at zero velocity, $[P(v)-G(\sigma_{\ast})]/P(v=0)$. The Gaussian
has the width given by the best fit to the distribution, $\sigma_{\ast}$, which
would be the observationally measured value.
The anisotropy radii considered are $r_a/s\rightarrow \infty$
(i.e., isotropic), $10$, $3$, and $1$, roughly spanning the range 
used in previous cosmological analyses.  The distributions are calculated using a Monte
Carlo sampling with $10^9$ particles and there is some residual shot noise. We used rectangular or circular extraction apertures from Table~\ref{tab:data}, as appropriate, after convolving the distributions with their corresponding PSF, modeled as a Gaussian. 

SDSS~J1206$+$4332 (Figure~\ref{fig:DF-1206}) has the largest negative deviations between the two distributions, leading to some of 
the largest underestimates of the
true rms velocity. The fractional differences between the measured $\sigma_{\ast}$ and $v_\text{rms}$ are $\sigma_{\ast}^2/v_\text{rms}^2-1 = 1.8\%$, $1.0\%$, $-2.6\%$ and $-7.0\%$
as we reduce $r_a$ and $h_4= -0.006$, $-0.005$, $0.004$ and
$0.018$ switches from slightly negative to positive as the anisotropy radius decreases in each model.
WFI~2033$-$4723 (Figure~\ref{fig:DF-2033}) has the 
strongest positive differences, with $\sigma_{\ast}^2/v_\text{rms}^2-1 = 
5.6\%$, $5.6\%$, $4.7\%$ and $0.1\%$ with $h_4= -0.015$, $-0.015$, 
$-0.013$ and $-0.002$. The differences probably arise because 
SDSS~J1206$+$4332 is a small lens ($s=0\farcs147$) observed in a 
large aperture ($0\farcs55 \times 0\farcs55$) so that significant 
changes in the anisotropy model can occur inside the aperture which is translated 
into a switch of sign in $\sigma_{\ast}^2/v_\text{rms}^2-1$ and in $h_4$. On 
the other hand, WFI~2033$-$4723 is a large lens ($s=0\farcs987$) compared to its aperture so the effects of the anisotropies at larger radii have less effect. 

The isotropic model is a particular case that is also covered by the choice of the constant anisotropy range by \cite{TDCOSMO2025}. For this case, the differences between the measured velocity dispersion and the root mean square velocity are $4.5\%$, $5.5\%$, $2.8\%$, $3.1\%$, $3.3\%$, $1.8\%$, $3.4\%$, $5.6\%$, and $3.7\%$ for DES~J0408$-$5354, HE~0435$-$1223, PG~1115+080, RX~J1131$-$1231 observed with NIRSpec, RX~J1131$-$1231 observed with KCWI, SDSS~1206+4332, B1608+656 , WFI~2033$-$4723, and WGD 2038$-$4008, respectively. For $r_a/s=$ 10, 3, and 1, the averages between the measured velocity dispersion and true rms are $3.4\%$, $1.1\%$, and $-2.5\%$, respectively. In each case, the lenses defining the lower and upper limits are SDSS~J1206$+$4332 and WFI~2033$-$4723, which is why there are used for Figures~\ref{fig:DF-1206} and \ref{fig:DF-2033}.

In summary, the differences between the measured velocity
dispersion and the mean square velocity are significant and can easily produce 
systematic corrections twice as big as the required uncertainty on $H_0$. Since we expect early-type galaxies
to be moderately radially anisotropic, the sense of the
correction is for the true mean square velocity to be larger
than the measured dispersion. If the velocity 
dispersions are not properly corrected, the models are less centrally
concentrated and, hence, lower values of $H_0$ are inferred.
 
\section{Models of the Anisotropy}
\label{sec:isotropy}

There is a puzzle in the results of \S\ref{sec:disp},
namely that the models had $h_4 < 0.018$ (and mostly $<-0.008$), while 
typical early-type local galaxies have $h_4 \simeq 0.01$ to $0.04$ (e.g., \citealt{Emsellem2004},
\citealt{Arnold2014},
\citealt{Veale2017}). There is evidence of a slight redshift evolution with medians of $h_4=0.019\pm0.002$ at $z = 0.82$ and $0.045\pm0.008$ at $z = 0.32$ (\citealt{DEugenio2023}) but there are not sufficiently strong to explain the range of values found in \S\ref{sec:disp}.  This has
two important implications.  First, the corrections we found
are likely significant underestimates.  For the rough scaling
of \cite{vanderMarel1993}, the fractional differences should
be $v_\text{rms}^2/\sigma_{\ast}^2 -1 \simeq 5 h_4$ and we should expect 
corrections of 5-20\%, not 1-5\%.  Second, the fact that
the O-M DFs are not producing sufficiently large $h_4$
values implies that they do not span the range of
physical properties occupied by early-type galaxies. 
In fact, there is a strong case that the O-M models 
are unlikely to be correct at either large or small
radii, which is one reason why it is not used in the most recent time-delay cosmography studies. But constant anisotropy models are not physical realizations of real galaxies and halos either.

The O-M models are forced to be isotropic at
their centers and radial in their outskirts with a fairly
steep transition due to the dependence on $r^2$. \cite{Mamon2006}
strongly argue that the Osipkov-Merritt models are a poor
representation of the anisotropy profiles found in simulations
and instead argue for 
\begin{equation}
   \beta = \frac{ 1}{ 2 } \frac{ r }{ r + r_a }
    \label{eqn:df2}
\end{equation}with $r_a \simeq 1.4 R_e \simeq 2.5 s$. This profile is
again isotropic at the center, but then slowly transitions
(because it scales with $r$ not $r^2$) to moderately
radial orbits ($\beta \rightarrow 1/2$ instead of unity) 
at large radii.

Physically, it is essentially impossible to have a halo 
comprised of nearly radial orbits.  A Hernquist
O-M model in isolation is unstable for
$r_a/s \ltorder 1.1$ (see, e.g., \citealt{Meza1997}),
and is unphysical for $r_a/s \ltorder 0.2$ (see, e.g., \citealt{Baes2002}).
In a cosmological scenario with early-type galaxies forming  
through mergers, the ``stirring'' produced by the mergers will
drive the halo to be comprised of moderately radial orbits
like the model in Eqn.~\ref{eqn:df2} (see, e.g., \citealt{Dekel2005}, \citealt{Onorbe2007}, 
\citealt{Wojtak2013}).  While it is straight
forward to compute velocity dispersion profiles based on
this model for the anisotropy, there is no easy way to check that
it corresponds to a physical DF (other than that it was found
to be a good fit to simulations) or to check whether it 
produces larger values of $h_4$.

There is also no good reason to think that the central regions
are truly isotropic.  The examples shown in \cite{Mamon2006}
are moderately more radial and less isotropic than a fit
using Eqn.~\ref{eqn:df2} at the smallest radii they show.
\cite{Gerhard2001}, in a sample of 21 early-type galaxies,
found that they almost all had moderately radial anisotropies
($\beta \sim 0.1$ to $0.3$) even at $0.2 R_e \simeq 0.4 s$.
\cite{Koopmans2009} use models of lenses with dynamical data
to argue for a moderate mean anisotropy $\langle \beta \rangle \simeq 0.45\pm0.25$
using constant anisotropy models.  To the extent that the
stellar dynamical measurement is dominated by the central
regions of the galaxy ($r<R_E$), this also argues for 
having moderately radial rather than isotropic orbits in
the inner regions.    

Where the O-M DF had the form $f(Q)$,
\cite{Cuddeford1991} showed that a DF of the form
$f(Q) L^{-2\beta_0}$ produces an anisotropy profile of 
\begin{equation}
    \beta(r) = \frac{ r^2 + \beta_0 r_a^2}{ r^2 + r_a^2 }.
    \label{eqn:df3}
\end{equation}Like the O-M DF, the orbits become increasingly radial in
the halo, but the central regions have an anisotropy of
$\beta_0$ rather than being isotropic.
\cite{Baes2002} show that for a Hernquist mass distribution,
there is a value of $r_a$ below which the models become
unphysical, at $r_a/s \ltorder 0.2$ for $\beta_0 = 0$
with slightly smaller (larger) limits for moderately 
radial (tangential) anisotropies. 

\begin{figure}
\centering
\includegraphics[width=0.47\textwidth]{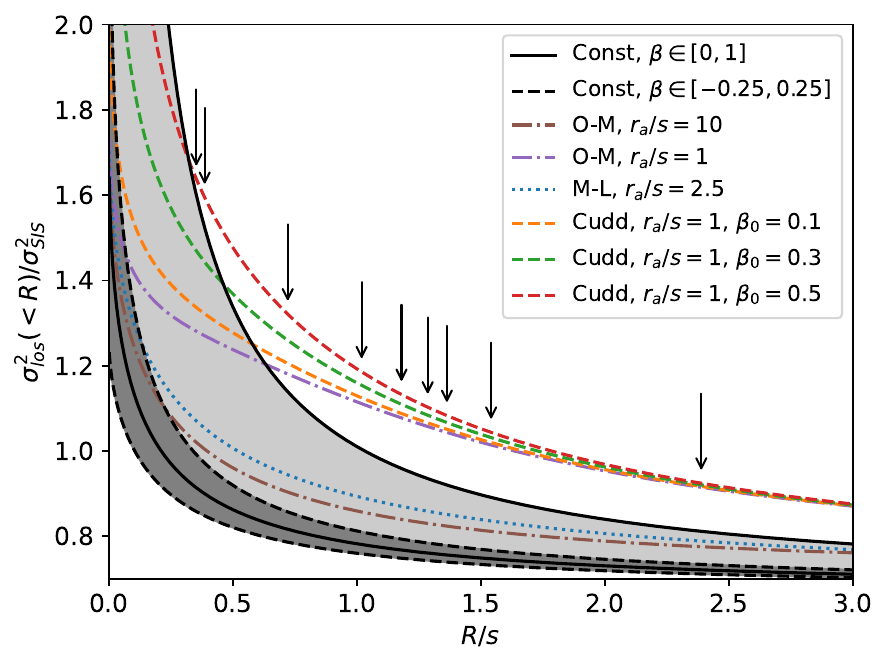}
\caption{ 
Mean squared los velocity dispersion inside radius $R$, $\sigma_{los}^2(<R)$, for the Hernquist profile embedded in an SIS mass model normalized by the squared velocity dispersion of the SIS model, $\sigma_{SIS}^2$. The solid lines are for the constant anisotropy model with $\beta=0$ (bottom) and $\beta=1$ (top) and the region between them is shaded in gray. The darker gray region between the black dashed lines is the anisotropy range used by \cite{TDCOSMO2025}. The dashdotted lines correspond to the O-M model with $r_a/s=10$ (bottom brown) and $r_a/s=1$ (top purple). The blue dotted line is for the anisotropy profile given in Eqn.~\ref{eqn:df2} with $r_a/s=2.5$. The dashed lines are for the anisotropy profile of Eqn.~\ref{eqn:df3} with $r_a/s=1$ and $\beta_0=0.1$ (lower orange), $0.3$ (middle green) and $0.5$ (upper red). The arrows are are placed at the effective circular aperture radius of the eight lenses in the order from left to right of WFI~2033$-$4723, HE~0435$-$1223, DES~J0408$-$5354, RX~J1131$-$1231 observed with NIRSpec, B1608$+$656, WGD~2038$-$4008, RX~J1131$-$1231 observed with KCWI, PG~1115$+$080, and SDSS~J1206$+$4332.
  }
\label{fig:isotropy}
\end{figure}

Whether changes in the anisotropy model matter for cosmological inferences depends on whether
they expand the range of velocity dispersions 
that can be produced at fixed mass. Figure~\ref{fig:isotropy}
shows the mean square los velocity dispersion inside radius $R$,
$\sigma_{los}^2(<R)/\sigma_{SIS}^2$, for a Hernquist stellar 
profile embedded in an SIS mass model for a constant anisotropy,  
the O-M model (Eqn~\ref{eqn:OM}) and the anisotropy profiles given in Eqns.~\ref{eqn:df2}
and \ref{eqn:df3}. The arrows are placed at the 
radius of the effective circular aperture of each lens taking into account 
the broadening due to seeing FWHM ($\left[ (\Delta x + 
2\sigma)(\Delta y + 2\sigma)/\pi\right]^{1/2}$ with $\sigma = 
\text{FWHM}/(8\ln 2)^{1/2}$). From left to right the arrows correspond to WFI 2033$-$4723, HE 0435$-$1223, DES J0408$-$5354, RX J1131$-$1231 observed with NIRSpec, B1608$+$656, WGD 2038$-$4008, RX J1131$-$1231 observed with KCWI, PG 1115$+$080, and SDSS J1206$+$4332. For the O-M model we show the cases
with $r_s/s=10$ and $r/s=1$, which roughly
corresponds to the range of the models typically used in previous time-delay models. The O-M model with $r_a/s=10$ lies inside the envelope of the constant anisotropy model if the most extreme case $\beta=1$ is considered. On the other hand, the $r_a/s=1$ model has higher velocity dispersions for radii larger than $\simeq0.65s$. Hence, the range of the O-M models can allow for larger velocity dispersions than constant anisotropy models.
We show the results for Eqn.~\ref{eqn:df2} with $r_a/s=2.5$,
and it basically lies inside the constant anisotropy models. Hence, using this anisotropy profile adds no more freedom
than is already provided by the current models.

Figure~\ref{fig:isotropy} also shows the dispersion
profiles for the Cuddeford anisotropy profile of Eqn.~\ref{eqn:df3} with 
$r_a/s=1$ and $\beta_0 = 0.1$, $0.3$ and $0.5$.  This model will be the same as the $r_a/s=1$ O-M model
for $\beta_0=0$.  These models expand the range of 
velocity dispersions in the sense of allowing higher velocity dispersions for the same dark
matter distribution. Although the \cite{Cuddeford1991} model with $r_a/s=10$ is not shown in Figure \ref{fig:isotropy} to clearly show the differences for $r_a/s=1$, it does not expand the velocity dispersion range towards lower values than the isotropic model. The differences between \cite{Cuddeford1991} models with $\beta_0=0.5$ and constant $\beta=1$ anisotropy models are large for 
velocity dispersions measured in apertures with intermediate to large
$R/s$. For example, the fractional change $\Delta\sigma_{\text{los,Cudd}}^2/\sigma_{\text{los,const}}^2$ for DES J0408$-$5354 is $+16\%$, for RX J1131$-$1231 observed with NIRSpec, B1608$+$656, WGD 2038$-$4008, and for RX J1131$-$1231 observed with KCWI the shift is roughly $+18\%$ and for PG 1115$+$080 and SDSS J1206$+$4332 the fractional changes slightly lower to $+17\%$ and $+14\%$, respectively. For the systems with smallest spectroscopic apertures, WFI 2033$-$4723 and HE 0435$-$1223, the differences are $+2\%$ and $+5\%$, respectively.

The anisotropy range used by \cite{TDCOSMO2025} is also shown in Figure~\ref{fig:isotropy}. Given the narrow range of constant anisotropies allowed, $\beta\in[-0.25,0.25]$, the range of velocity dispersions that the current models can produce is limited and, if the actual anisotropy distribution produces results outside the range allowed by the model, then the $H_0$ inference will be biased. If the true distribution corresponds to one of the explored models, a constant anisotropy model would need to reduce the amount of dark matter to make the mass distribution more concentrated, which would drive the value of $H_0$ upwards. The fractional changes in the velocity dispersion of different anisotropy models with respect to the maximum range that the constant anisotropy model can achieve are presented in Table~\ref{tab:summary}. We should note here that we are discussing changes in the model dispersions not in the measured velocity dispersion, so that fractional changes in the maximum allowed velocity dispersion will drive $H_0$ upwards but it may not directly imply the same fractional change. Including models with moderately radial orbits in the central regions such as \cite{Cuddeford1991} or constant anisotropy models
is more physical but they will also produce larger values of
$h_4$. This will lead to larger corrections between 
$\sigma_*$ and $v_\text{rms}$ which will counteract the allowed higher central velocity dispersions
of centrally anisotropic models.

\section{The Photometric Model of the Galaxy}
\label{sec:profile}

\begin{figure*}
\centering
\includegraphics[width=\textwidth]{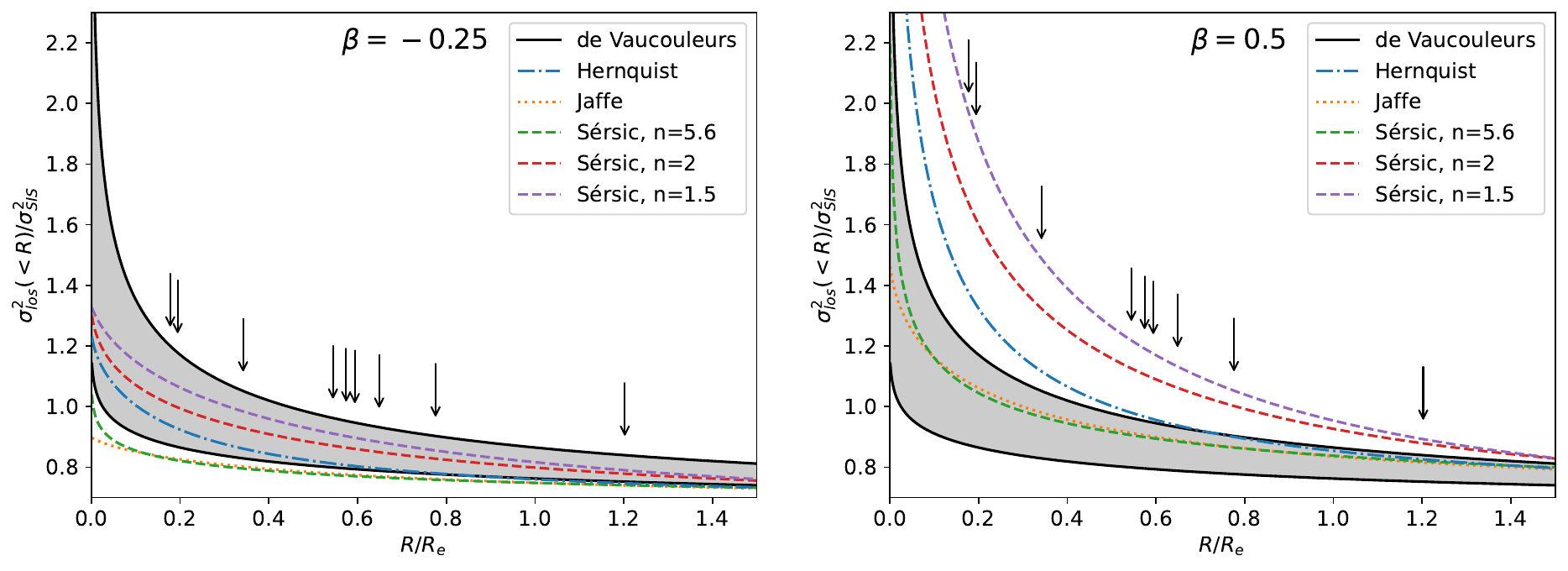}
\caption{
Mean squared enclosed los velocity dispersion, 
$\sigma_{los}^2(<R)/\sigma_{SIS}^2$, for various stellar
density distributions normalized to the same effective radius $R_e$ embedded in an SIS mass distribution and using a constant anisotropy model. The solid lines and the shaded region between them are the $\sigma_{los}^2(<R)/\sigma_{SIS}^2$ that the \protect\cite{deVaucouleurs1948} profile (or equivalently \protect\cite{Sersic1968} 
profile with $n=4$) can produce, the blue dashdotted 
lines are for the \protect\cite{Hernquist1990} profile and the orange dotted lines are for the \protect\cite{Jaffe1983} profile. The 
dashed lines are for the \protect\cite{Sersic1968} 
profiles with $n=5.6$ 
(green), $n=2$ (red) and $n=1.5$ (purple). The models on the left (right) panel have an anisotropy of $\beta=-0.25$ ($\beta=0.5$) except for the \protect\cite{deVaucouleurs1948} profile where both anisotropy profiles are shown for reference. The 
arrows are located at the approximate value of $R/R_e$ 
corresponding to the effective aperture area for each lens and have the same ordering as in Figure \ref{fig:isotropy}.
  }
\label{fig:profiles}
\end{figure*}

We summarized the choices of photometric models in Sect.~\ref{sec:obs}. Here we discuss the consequences of errors in photometric models and the general problem of color gradients. Some consequences for earlier models are discussed in Appendix~\ref{app:prev}. While we are generally staying within the assumption
of spherical galaxies, there is the question of what
scale length of an elliptical galaxy should be used
in a spherical analysis. Three standard possibilities
are the major, intermediate, and minor scale lengths,
whose ratios of $1:q^{1/2}:q$ are determined by the 
axis ratio $q$. 
New TDCOSMO models explicitly state that intermediate axis was used (e.g., \citealt{Shajib2020,Shajib2023,TDCOSMO2025}).
However, of the six lenses analyzed by H0LiCOW, the
model for the light distribution is only fully specified
in \cite{Suyu2013}.  In this study of RX~J1131$-$1231,
the effective radius is clearly defined as the major 
axis effective radius (their Eqn.~15) and then used
in the dynamical analysis. We would argue that the
intermediate axis seems the most physical choice. 

We can ask how the velocity dispersions for a 
Hernquist in Hernquist model (Eqn.~\ref{eqn:herndisp}
at fixed M) or Hernquist in SIS model (Eqn.~\ref{eqn:sisdisp}
at fixed $\sigma^2$) change given a change in the
scale length from the major axis to the intermediate axis. The fractional changes for $r \sim s$
are $\sim -\Delta s/s$ and $\sim + \Delta s/2 s $,
respectively.  In the first case, reducing $s$ increases
$\sigma_H^2$ because both the stellar and mass distributions
are becoming more compact. In the second case, reducing
$s$ decreases $\sigma_S^2$ because the mass distribution is
fixed even as the stellar distribution becomes more compact.
In a more realistic model, combining the stars with a
dark halo, shrinking the scale length should 
generically increase the velocity dispersion because
the stars dominate the mass of the central regions of
galaxies.  This should mean that modeling a lens with
an overestimate of the scale radius underestimates the
velocity dispersion for a given mass inside the Einstein
ring, driving the models to less dark matter and higher
estimates of $H_0$.

Figure~\ref{fig:profiles} shows $\sigma^2_{los}(<R)/\sigma_{SIS}^2$
using the SIS mass model for the current and previous 3D stellar mass choices:  
Hernquist, Jaffe, and $n=1.5$, $2$, $4$ (i.e., de Vaucouleurs) and $5.6$ S{\'e}rsic
models, holding the
effective radius $R_e$ fixed.  We show
the constant anisotropy models for $\beta=-0.25$ and $\beta=0.5$ (left and right panel, respectively) to encompass the \cite{TDCOSMO2025} anisotropy ranges and other anisotropy measurements in early-type galaxies \citep[e.g.][]{Gerhard2001,Koopmans2009}.
We also roughly indicate the aperture radius plus seeing broadening corresponding
to each lens. A mismatch between the assumed stellar density profile and the current mass distribution will only be important if the actual mass profile allows 
dispersions that lie outside the envelope spanned by the assumed mass model.

\cite{TDCOSMO2025} fit the light profiles of the time-delay lenses with single or multiple \cite{Sersic1968} profiles with the exception of B1608+656 that is fitted with a \cite{Hernquist1990} profile. We compare the velocity dispersion that the S\'ersic profile fit (or the dominant component in the case of double and triple S\'ersic fits) produces with respect to a S\'ersic profile with an index 20\% larger or smaller to account for possible uncertainties in the light model. For the case of B1608+656, we compute the differences with respect to S\'ersic profiles of indices $n=3.2$ and 4.8 since the ranges of the Hernquist velocity dispersions at the size of its aperture (fifth arrow from the left in Figure~\ref{fig:profiles}) are similar to a de Vaucoulerus profile.

Since the dispersion curves for the \cite{Sersic1968} models steadily lie at higher dispersions as $n$ decreases, overestimating the profile index will produce a model with a lower range of velocity dispersions whereas underestimating the index will produce higher velocity dispersions. If the actual mass profile has a lower (higher) S\'ersic index than the adopted model, the dynamical model will need to decrease (increase) the dark matter to accommodate the measured velocity dispersion, which will drive the value of $H_0$ upwards (downwards). The fractional changes on the velocity dispersion for each lens due to modifying the mass profile index by 20\% can be found in Table~\ref{tab:summary}. The changes are larger for the lenses measured in smaller apertures since the model velocity dispersions tend to diverge more rapidly at smaller radii. Hence, the impact of the uncertainties in the photometric model may become more important in higher resolution observations.

Early-type galaxies also have color gradients due to some combination of age and metallicity gradients (see, e.g., \citealt{Peletier1990}, \citealt{LaBarbera2005}, \citealt{Kennedy2016}), which implies the presence of multiple stellar populations with different kinematics. Consider the spherical, isotropic Jeans equation
\begin{equation} 
   \frac{1}{n } \frac{ d n \sigma^2}{dr}
     = -\phi'
      \label{eqn:sphjeans}
\end{equation}where $\sigma$ is the velocity dispersion, 
$n$ is the stellar number density and $\phi'$ is the radial derivative of the potential.  Stellar dynamical calculations generically assume that $n \propto  j$ where $j$ is the deprojected photometric profile in some band and the constant of proportionality cancels in Eqn.~\ref{eqn:sphjeans}.  In the presence of color/age/metallicity gradients (and, therefore, different stellar populations), however, different choices for defining $n$ will lead to different inferences about the mass.  For example, suppose we have two power law distributions $n \propto r^{-\alpha}$ and $n\propto r^{-\beta}$.  The inferred mass distributions will have fractional differences of order $|\Delta M/M| \sim |\alpha-\beta|$ and the galaxy will have a color gradient of
$d (m_\alpha-m_\beta)/d\log r = 2.5(\alpha-\beta)$.  Since typical color gradients
are $0.1$-$0.2$~mag/dex (e.g, \citealt{Peletier1990}), 
they can lead to 4-8\% errors in the inferred mass.  

This leads to the problem of determining what to use for $n$.  A simple model problem
gives the basic answer.  Assume the stellar populations can be divided into populations
characterized by $n_i$ and $\sigma_i$ each of which satisfies Eqn.~\ref{eqn:sphjeans}.
Consider a spectral region with a single Gaussian absorption line where each population makes a flux 
contribution of $\ell_i$ with a line equivalent width $w_i$.  Their combined 
spectrum in velocity space is
\begin{equation}
     s(v) = \sum_i \ell_i n_i \left( 1 - \frac{w_i}{\sqrt{2\pi}\sigma_i}
        \exp\left( -\frac{1}{2}\frac{v^2}{\sigma_i^2}\right)\right).
\end{equation}If we subtract the continuum $c = \sum_i l_i n_i$ we are left with the weighted
contribution of the populations to the absorption line.  The normal ways
of modeling this essentially treat this as the velocity probability distribution
where we would measure a mean square velocity
\begin{equation}
      \langle v^2 \rangle = \frac{ \sum_i \ell_i n_i w_i \sigma_i^2}{\sum_i \ell_i n_i w_i}
\end{equation}and we need
\begin{equation} 
   \frac{1}{n } \frac{ d n \langle v^2\rangle}{dr}
     = -\phi',
      \label{eqn:sphjeans2}
\end{equation}which is only true if 
\begin{equation}
    n \propto \sum_i \ell_i n_i w_i.
    \label{eqn:weighting}
\end{equation}The Jeans equation for the combination of the populations must use an $n$ which is proportional to the weighting of the populations in the determination of the mean square velocity. While \cite{TDCOSMO2025} use the light profile in the wavelength range of the absorption lines used to measure the velocity dispersion, this is not the same as the weighting required by Eqn.~\ref{eqn:weighting} and so does not mitigate this systematic problem. To our knowledge this issue has never been properly taken into account.

\section{Early-Type Galaxies as a Homogeneous Population}
\label{sec:homogeneous}

There is a great deal of evidence that early-type galaxies
are a fairly homogeneous population, perhaps with some 
mass-dependent trends.  A considerable portion of the
evidence for this homogeneity, particularly outside the
local universe, comes from studies of gravitational lenses.
Homogeneity has important consequences for the 
analysis of the lens sample. Most of the lenses selected for cosmological studies are single\footnote{Note, however, the exception of B1608$+$656, where the lens is clearly not a single component early-type galaxy but
an interacting pair of galaxies inside the Einstein ring.  It is
highly unlikely that the spherical Jeans equations are appropriate
for any analysis of the dynamics of this system.} early-type galaxies. Treating the model choices independently for each lens and combining the results afterwards has an impact on the uncertainties recovered for $H_0$. Analyses like \cite{TDCOSMO2025} take some of these issues into account, but it is useful to review the broader issues.

As an example, \cite{Wong2020} treat the anisotropy as if 
it is an independent variable for each lens galaxy.
In general, the available data cannot determine the
anisotropy radius, so the probability of a given $H_0$ 
can be modeled as
\begin{equation}
    \frac{d^2 P}{dH d r_a }\propto \exp\left( - \frac{1}{2 } \frac{ \left( H + \alpha r_a \right)^2}{\sigma_H^2} \right).
\end{equation}For a fixed anisotropy radius, the uncertainty is $\sigma_H$, and 
there is a degeneracy with changes in the anisotropy radius
$r_a$ characterized by the coefficient $\alpha$.  We have centered the distributions
on zero to simplify the expressions.  Although H0LiCOW uses a uniform
prior, we will assume that $r_a$ is constrained
by a Gaussian of width $\sigma_a$, 
$P(r_a) \propto \exp(- r_a/2\sigma_a^2)$,
so that our calculations are analytic. 

If the anisotropy radius is assumed to be a random variable for each lens, the joint probability of $H_0$ from $N$ lenses has the form
\begin{equation}
  \frac{ d P}{ dH } = \int \Pi_{i=1}^N dr_{a,i} P(r_{a,i})
     \frac{ d^2 P}{dH dr_{a,i}}. 
\end{equation}where the final result is marginalized over the $N$ independent 
isotropy radii $ r_{a,i}$.
For our simplified Gaussian model, this leads to a mean square 
error in $H_0$ of 
\begin{equation}
    \frac{ \sigma_H^2}{N }+ \frac{ \alpha^2 \sigma_a^2 }{N}.
\end{equation}By assuming that early-type galaxies are a completely heterogeneous
dynamical population, the contributions of both the random errors characterized by 
$\sigma_H$ and the systematic errors from the anisotropy decrease 
as $N^{-1/2}$.  However, if the lenses are a dynamically
homogeneous population, then a single
value of $r_a/s$ characterizes all the lenses in the sample.
In this case, the marginalized probability is 
\begin{equation}
  \frac{ d P }{ dH } = \int dr_a P(r_a) 
   \Pi_{i=1}^N \frac{ d^2 P_i }{dH dr_a}.
\end{equation}and the mean square error is 
\begin{equation}
    \frac{ \sigma_H^2 }{ N} + \alpha^2 \sigma_a^2.  
\end{equation}
For a dynamically homogeneous population, 
the contribution of the uncertainties in the anisotropy 
to the uncertainties in the Hubble constant is completely unchanged 
by the number of lenses used in the analysis. A clear example can be seen in \citet{Birrer2021}, where the predicted precision of $H_0$ with a sample of 40 lenses is not reduced significantly compared to the precision achieved with 7 lenses when some parameters are treated on the population level as described in \cite{Birrer2020}.

This argument also
holds for any choice for the stellar density profiles used in the
dynamical analysis unless they are specific fits to individual 
lenses as in \cite{TDCOSMO2025}. The issue still holds for any parameterization of the dark matter density distribution. For example, the break radius of the dark matter distribution as compared to the effective radius of the stellar mass distribution.  For example,  \cite{Suyu2010} and \cite{Wong2017} experimented with using either
the Hernquist or Jaffe models for the stellar dynamical models
finding ``small'' or $\sim 1\%$ offsets between the two models.   Viewed as an uncertainty to be marginalized over, one should
really model all the lenses as Hernquist (Jaffe) to get the probability
distribution of $H_0$ assuming the Hernquist (Jaffe) model and
then marginalize over the two model possibilities.  One should
not marginalize the individual results over the two profiles
and then combine the lenses -- this corresponds to the assumption
that each lens should be randomly assigned to have one or the other profile.
As a homogeneous population, they should all be assigned to one 
or the other profile. Since the results no longer statistically
average the uncertainties created by the profiles over the lenses,
even ``small'' differences become increasingly important as the
overall uncertainty shrinks.   

\begin{figure}
\centering
\includegraphics[width=0.50\textwidth]{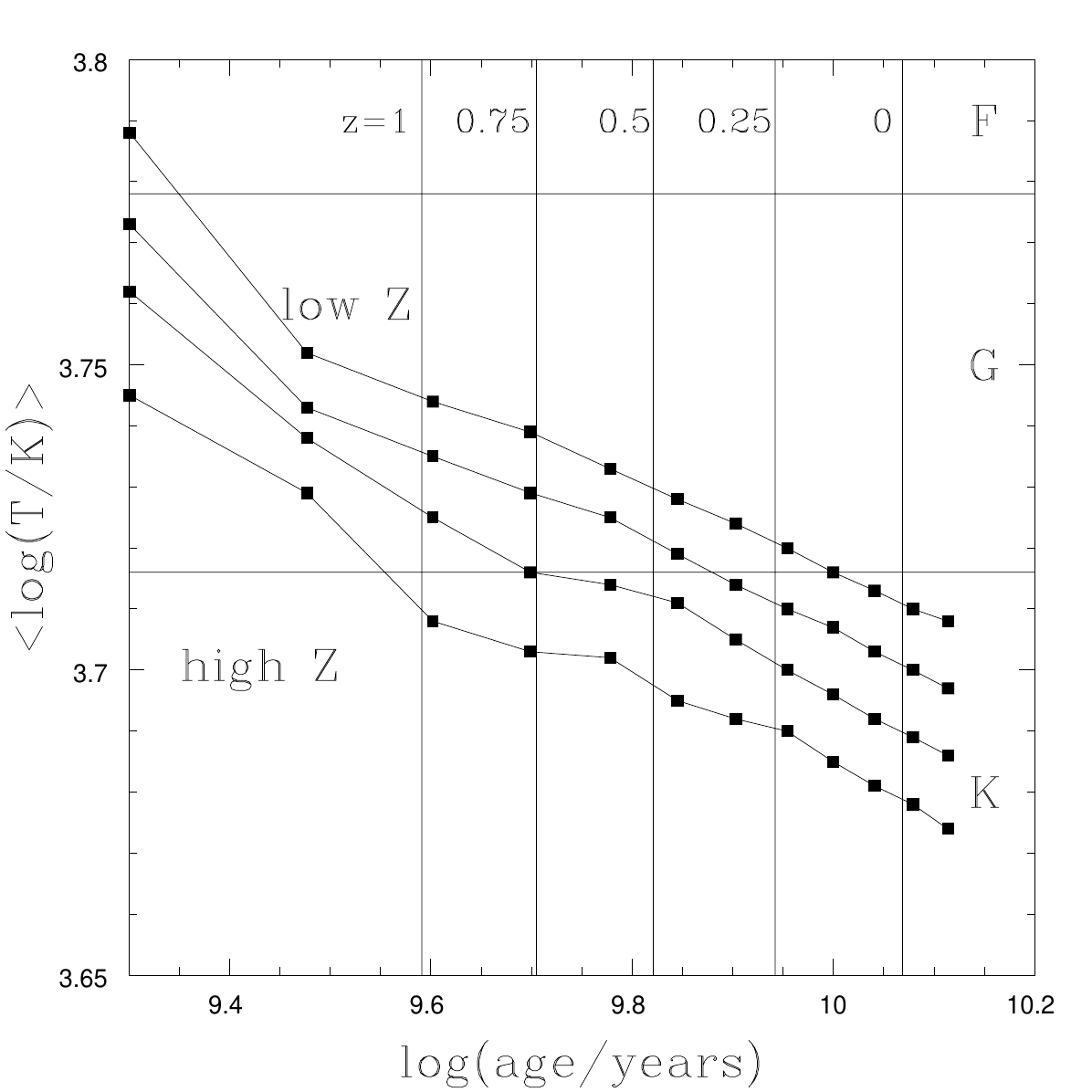}
\caption{ B-band luminosity weighted mean temperatures
$\langle \log T \rangle$ as a function of age and 
metallicity for the standard PARSEC isochrones. The
metallicities are roughly twice solar (lowest), 
solar, half solar and quarter solar (top).  Vertical
lines indicate the redshift corresponding to that 
age if the stars form 2~Gyr after the big bang.  
Horizontal lines roughly indicate the spectral types
associated with the temperatures.
  }
\label{fig:temps}
\end{figure}

The  stellar populations of early-type galaxies are also very homogeneous
between galaxies of similar velocity dispersion (e.g., \citealt{Bernardi2003},
\citealt{Kuntschner2006}) aside from a steady evolution in redshift
(e.g., \citealt{Rusin2005}, \citealt{Treu2006}).  This means
that there is a similar issue associated with the dependence of 
the velocity dispersion measurements on the stellar template used in the analysis. Both
H0LiCOW and TDCOSMO treat the contribution of the stellar templates to the 
velocity dispersion uncertainties as an additional contribution to 
the random uncertainties with no correlations between lenses. \cite{TDCOSMO2025} use the recipe from \cite{Knabel2025} to try to reduce the systematic errors arising from the stellar templates. However, any residual systematic errors are treated as independent errors for each individual lens.

Here we present a simple case to exemplify how stellar templates produce systematic biases at the population level. Figure~\ref{fig:temps} shows the B-band luminosity 
weighted mean stellar temperature ($\langle \log T\rangle$) for the
standard PARSEC model isochrones as a function of age for metallicities
of $Z=0.0052$, $0.010$, $0.021$ and $0.041$ (roughly $1/4$, $1/2$,
$1$ and $2$ times solar).  We weight by the B-band luminosity 
because this is roughly the rest wavelength range usually modeled to 
determine the velocity dispersion.  The mean surface gravity is
not very age-dependent, but does increase with metallicity. The
solar/super-solar models have $\langle \log g \rangle \simeq 3.4$,
while the lower metallicity models have $\langle \log g \rangle \simeq 3.2$
and $3.1$.  This may seem to conflict with the idea that red giants
dominate the emission from early-type galaxies, but it is due to the weighting by
the B-band luminosity.  Weighted by the I-band luminosity of the stars,
the mean stellar temperatures are significantly lower and the surface
gravities are lower and typical of red clump stars. 

If we assume that the spectra of early-type galaxies depend only on age
(redshift) and velocity dispersion, then the pattern of differences in
dispersion estimates using different stellar templates will be systematic,
not random.  If there is a 5\% difference in $\sigma_*$ from using star
A instead of star B for one early-type galaxy, it should show the same
offset for another early-type galaxy of similar redshift and age and a
systematic pattern of offsets as a function of age and dispersion.  
This again means that the uncertainties created by choices of template
stars cannot be viewed as a random uncertainty associated with each 
lens.  Instead, a probability distribution for $H_0$ should be computed
from all lenses for each template star (weighted by the goodness of fit of the template as suggested by \citealt{Knabel2025}), and then the final result should
be marginalized over the choices of template stars.

There are four stars in the INDO-US \citep{Valdes2004} sample that have temperatures
and surface gravities similar to these means and with roughly Solar abundance
or higher (G16$-$32 (K1IV) with $T=4768$~K,
$\log g = 3.40$ and $[Fe/H]=-0.07$; HD197964 (K1IV) with $T=4990$~K, $\log g=3.49$,
and $[Fe/H]=0.13$; HD92588 (K1IV) with $T=5044$, $\log g= 3.60$ and $[Fe/H]=-0.10$;
and HD191026 (K0IV) with $T=5150$~K, $\log g = 3.49$ and $[Fe/H]=-0.10$). We
would expect that stars with different physical properties would lead to 
different velocity dispersions.   
As a first experiment, we took a one in ten random sparse sampling of the 
\cite{Valdes2004} G and K stars with temperatures between
3800~K and 5600~K and cross fit them, modeling star A by star B
convolved with a velocity-shifted Gaussian plus a fifth order polynomial and the
reverse, keeping the best fit in the combined results.  The scatter
in the dispersions of $63$~km/s is well-modeled by 
\begin{equation}
    \sigma_T = a \Delta T + b \Delta\log g + c \Delta [Fe/H]
    \label{eqn:scatter}
\end{equation}where $a = 0.061 \pm 0.001$, $b = -5.5 \pm 0.6$ and $c=-48.5 \pm 0.7$ and with a
residual scatter of $20$~km/s.  Fitting a star with a higher temperature
using one with a lower temperature ($\Delta T > 0$) requires a higher 
dispersion. Fitting a star with a lower $\log g$ using one with a higher
surface gravity ($\Delta \log g < 0$) requires a higher dispersion.  
And, finally, fitting a star with a lower $[Fe/H]$ using one with a higher
metal abundance ($\Delta [Fe/H] < 0$) requires a larger dispersion.  The
first correlation clearly arises because the hotter stars have broader
lines.  The abundance correlation probably arises not because the lines
are broader for lower abundance but because smearing the stronger lines 
of the more metal rich star provides the best fit for matching the weaker 
lines of the lower metallicity star.  

Next we repeated the process after first convolving the spectrum being
fit with a $\sigma_* = 250$~km/s Gaussian.  Naively, if star A
was fit by star B convolved by $\sigma_T$ we might expect to measure
a dispersion of $\left(\sigma_*^2-\sigma_T^2\right)^{1/2}$, while if the
reverse held we might expect to measure $\left(\sigma_*^2+\sigma_T^2\right)^{1/2}$,
corresponding to simply combining the convolutions.  After removing
the worst fits, we found a scatter of about $30$~km/s and no significant
reduction in the scatter if we modeled the residuals as in 
Eqn.~\ref{eqn:scatter}.  This was moderately surprising and we
presently have no good explanation for the result.  

The effect of assigning a systematic
error to individual lenses and then assuming it is a random variable from 
lens to lens is that the net contribution to the error budget diminishes
with the number of lenses.  But for a homogeneous lens population, this assumption 
is not true -- if one lens galaxy has radially biased orbits, they 
almost certainly all have radially biased orbits, if one template star
produces a higher dispersion estimate than another template star for
one galaxy, it probably does so for all the lens galaxies, and so on.
The only question is the exact degree of correlation and covariance for all the population-level variables. The correlation and covariance of the measured velocity dispersion between galaxies (and between spatial bins) due to the stellar templates can be assessed and \cite{Knabel2025} find a positive correlation between early-type galaxies as expected. While the covariance is small, it is not negligible compared to the necessary tolerances and the effect on the population level will become increasingly important as the lens sample size increases and higher accuracies are needed.

\section{Discussion}
\label{sec:disc}

\begin{table*}
    \caption{Summary of the maximum percentage changes in the squared velocity dispersion, $\Delta\sigma^2/\sigma^2$, due to possible sources of systematic errors for the eight time-delay lenses.}
    \label{tab:summary}
    \centering
    \movetableright=-1in
    \resizebox{\textwidth}{!}{%
    \begin{tabular}{l SSSSSSSSS}
    \hline
    \hline
    \multirow{2}{*}{$\Delta\sigma^2/\sigma^2$ (\%)} & {DES~J0408}& {HE~0435} & {PG~1115} & {RX~J1131} & {RX~J1131} &  {SDSS~J1206} & {B1608} &  {WFI~2033} & {WGD~2038}\\
     & {(MUSE)} & {(NIRSpec)} & {(NIRSpec)} & {(NIRSpec)} & {(KCWI)} & {(NIRSpec)} & {(NIRSpec)} & {(NIRSpec)} & {(MUSE)} \\
    \hline
    \textsc{Point Spread Function} & & & & & & & & & \\
 Larger seeing & {$-$}0.7 & & & & {$-$}0.4& & & & {$-$}1.5\\
 Smaller seeing & {$+$}0.6 & & & & {$+$}0.3& & & & {$+$}1.5\\
 Miscentering & {$-$}0.3 & & & & {$-$}\hspace{-0.5em}0.03& & & & {$-$}\hspace{-0.5em}0.13\\
 PSF Single Moffat & {$-$}1.9 & & & & {$-$}1.3& & & & {$-$}2.0\\
 PSF Double Moffat & {$-$}4.5 & & & & & & & & {$-$}1.2\\
 \hline
    \textsc{Measured Dispersion} & & & & & & & & & \\
 Isotropic velocity DF & {$+$}4.5& {$+$}5.5& {$+$}2.8& {$+$}3.1& {$+$}3.3& {$+$}1.8& {$+$}3.4& {$+$}5.6& {$+$}3.7\\
 \hline
    \textsc{Anisotropy model} & & & & & & & & & \\
 O-M $r_a/s$=1 & {$+$}3.6& & {$+$}13.8& {$+$}10.6& {$+$}13.4& {$+$}12.7& {$+$}12.4& & {$+$}13.1\\
 Cuddeford $\beta_0$=0.1 & {$+$}5.8& & {$+$}14.4& {$+$}14.2& {$+$}12.0& {$+$}13.0& {$+$}13.4& & {$+$}14.0\\
 Cuddeford $\beta_0$=0.3 & {$+$}10.6& & {$+$}15.7& {$+$}14.9& {$+$}15.9& {$+$}13.4& {$+$}15.7& & {$+$}15.9\\
 Cuddeford $\beta_0$=0.5 & {$+$}16.0& {$+$}4.7& {$+$}17.1& {$+$}18.1& {$+$}17.7& {$+$}13.9& {$+$}18.1& {$+$}2.4& {$+$}17.9\\
 \hline
    \textsc{Photometric model} & & & & & & & & & \\
 S\'ersic $n=0.8n_\text{fit}$ & {$+$}5.6& {$+$}9.5& {$+$}3.2& {$+$}6.8& {$+$}5.5& {$+$}1.6& {$+$}3.0& {$+$}9.9& {$+$}5.9\\
 S\'ersic $n=1.2n_\text{fit}$ & {$-$}1.9&  {$-$}2.9& {$-$}1.4& {$-$}2.9& {$-$}2.5& {$-$}0.8& {$-$}2.9&  {$-$}3.0& {$-$}2.4\\
    \hline
    \end{tabular}}
\tablecomments{RX~J1131$-$1231 is repeated because it was observed with two different instruments.}
\end{table*}

A fundamental issue to consider for time-delay lenses are
the requirements for making stellar dynamical measurements with systematic 
uncertainties in $\sigma^2$ of order 2\% today and still smaller if
time delay cosmography is to be competitive as other methods improve.
Traditional uses of stellar dynamics to study dark matter really use ratios 
of the measurements, changes in the velocity dispersion with radius, which 
are less sensitive to calibration errors. Lensing compares mass estimates from the lens geometry to mass estimates from the kinematics and so are directly dependent on the absolute calibration of the velocity when constraining the mass distribution. 
Here we examined a broad range of systematic considerations involved in
combining stellar dynamical models with gravitational lensing. We summarize the fractional changes in the velocity dispersion in Table \ref{tab:summary} for the eight lenses we examined. Of the systematic issues we consider, only the line spread function was 
not presently a source of systematic errors on the order of the
required uncertainties. 

There are three issues with the PSF model.  First, for the FWHM of the
PSF must be known to better than with a 10\% accuracy if a precision of order
2\% is pursued.  If the FWHM is underestimated, models will be
driven to have more dark matter so as to reduce the central velocity
dispersion. This systematic is specially important for long-slit spectra of extended
objects (see Appendix~\ref{app:prev}), like most of the measurements used in time-delay cosmography prior to \citealt{TDCOSMO2025}. Long-slit spectra frequently lack a direct measurement of the seeing
corresponding to the spectroscopic observation.  Measures 
external to the instrument (e.g., differential image motion 
monitors, DIMMs, or wavefront sensors) are frequently measuring 
the seeing along different optical paths and do not include the contribution 
of the instrument. A FWHM measured from stars in an acquisition 
image will likely underestimate the FWHM in the spectroscopic
observation because of the accumulation of guiding errors
and seeing fluctuations during the far longer spectroscopic 
observation. Alternatively,
since most time delay lenses have high resolution HST observations,
directly modeling the lens profile along the slit may be
the best course although this is complicated by the
multiple source structure of lenses (lens plus images) and
the need to synthesize an HST-like imaging band pass from
the spectrum. A clear example on how this would impact the measured velocity dispersion can be found in \cite{Shajib2023} for RX~J1131$-$1231 where the luminosity-weighted velocity dispersion from the 2D velocity map within the extraction aperture used in \cite{Suyu2013} is 11\% smaller than the long-slit measurement.

The second issue is the possible shift between the center of the galaxy and the aperture. If the miscentering is small (i.e., by half the pixel size), this issue plays a minor role in the accuracy of the velocity dispersion. However, if precisions below 1\% in $H_0$ are pursued, then the lens galaxy must be precisely placed at the center. In any case, miscentering the galaxy drives the measurement of the velocity dispersion to lower values, underestimating the value of $H_0$.

The last measurement issue that enters is the assumed structure of the PSF.  Astronomical
PSFs have extended wings beyond a Gaussian core. If the true PSF is
actually the $\beta=4$ \cite{Moffat1969} model or the double Moffat model
of \cite{Racine1996} instead of a Gaussian,
the fractional corrections of the velocity dispersion can be as high as 2.0\% for the
single $\beta=4$ profile, and 4.5\% for the double 
Moffat model. In essence, ignoring the broader wings of the non-Gaussian
profiles is like underestimating the seeing and the models will
be driven to have less centrally concentrated mass distributions which leads to a bias in the $H_0$ inference towards lower values. Although most of the PSFs of time-delay lenses were usually modeled as Gaussians, more recent works like \cite{Rusu2020}, \cite{BuckleyGeer2020}, \cite{Mozumdar2023} and \cite{TDCOSMO2025}, for example, do use Moffat profile fits to the PSF in some cases. 

Some problems from PSF models can be reduced, but not eliminated, in space-based observations. Many newer observations (both ground- and space-based) are acquired with IFUs, possibly providing the PSF of a nearby uncontaminated star in the field to constrain both the profile shape and the FWHM may reduce the systematics, modulo the color dependence of the PSF. Given that over- and underestimating the FWHM has opposite effects on the $H_0$ inference, small misestimates of this quantity may cancel out on the population level. However, if the majority of the observations are acquired with the same instrument and the PSF is fitted with the same methodology, there will be residual systematic biases.

These observational effects are typically smaller than the potential issues
associated with the velocity distribution function, orbital anisotropy, and the photometric model. As discussed in \cite{vanderMarel1993}
and emphasized in \cite{Kochanek2006} for the problem of using stellar
dynamics in gravitational lens models, the standard observational quantity
called the velocity dispersion, $\sigma_{\ast}$, is not the root mean square velocity needed for the Jeans
equations, $v_\text{rms}$, unless the los velocity function is a Gaussian.
Since we expect stellar orbits to be biased to be radial rather than 
tangential, the sense of the correction is that the actual mean square
velocity for most of the lenses is larger than the measured dispersion. If this correction is not applied, the models are driven
to estimate lower values $H_0$. 

The O-M model we used to explore these issues has Gauss-Herminte coefficients $h_4\lesssim0.018$ which does not span the
range observed for early-type galaxies even at the redshifts of the lens galaxies.
The problem is likely due to the O-M model assumption that the centers of
galaxies are isotropic.  Real galaxies probably have a modest radial
anisotropy even in the central regions, which drives the larger values
of $h_4$ than are produced by the O-M models. This issue may be solved with the model choice in \cite{TDCOSMO2025}, where a constant anisotropy model is assigned to each component of the photometric profile. However, it would be interesting to study the general anisotropy profiles that arise when combining all components for each lens and study the $h_4$ created by these profiles to ensure that they span the observed range of early-type galaxies. Given that measurements of the fourth-order moment $h_4$ in early-type galaxies are in the order of 0.01 to 0.04 \citep{Emsellem2004,Arnold2014,Veale2017,DEugenio2023}, the correction between $v^2_\text{rms}$ and $\sigma_\ast^2$ would be on the order of 5-20\%, far above the required precision for $H_0$ inferences. Hence, it is crucial to include this correction when solving the Jeans equation independently of the choice of an anisotropy model.

We analyzed the impact of changing the assumed anisotropy model. The velocity dispersion profile ranges for the \cite{Mamon2006} anisotropy model and most of the range produced by the O-M model is spanned by constant anisotropy models with $0\leq\beta\leq1$. On the other hand, the \cite{Cuddeford1991} anisotropy model produces a broader range of velocity dispersions profiles than encompassed for these models, which will drive $H_0$ towards lower values. The differences are biggest for galaxies observed in apertures comparable to the photometric scale length or larger. The \cite{Cuddeford1991} anisotropy models have larger $h_4$ values, which will increase the
difference between the measured dispersion and the mean square velocity, shifting $H_0$ to the opposite direction. \cite{Liang2025ani} examined some of these issues for the constant anisotropy and O-M models reporting average differences in a mock sample rather than the expected shifts for specific lenses. Which anisotropy model is best can be constrained by spatially resolved kinematic data of nearby galaxies \citep[see, e.g.,][]{Bacon2001,Cappellari2011,Sanchez2012,Bryant2015,Bundy2015}, although it is still necessary to measure or estimate $h_4$ in order to correctly solve the Jeans equations.

In this work we focus on the impact of the anisotropy model choice when the central velocity dispersion is the only the kinematic constraint. There is, however, a strong observational effort to measure spatially resolved kinematics to break the mass-anisotropy degeneracy (see, e.g., \citealt{Shajib2023}, \citealt{TDCOSMO2025} or \citealt{Shajib2026} for the radial and 2D kinematics of RXJ~1131$-$1231; \citealt{Sheu2026} for the 2D kinematics of the time-delay lens SDSS~J1433+6007; or the 2D kinematic maps of the SLACS sample in \citealt{Knabel2026}). The prospects are that $H_0$ can be constrained with a precision better than $\le$4\% when spatially resolved kinematics constraints are included \citep[see, e.g.,][]{Birrer2021,Yildrim2020,Yildrim2023} and with a sub-percent bias \citep{Verma2026} provided the anisotropy model is a fair representation of the actual anisotropy of the lens galaxy. Even with measured 2D kinematics, the sense of the corrections will be the same that we find for the central velocity constrains (although they are expected to be smaller), and the correction between $\sigma_\ast$ and $v_\text{rms}$ with $h_4$ should still be applied. 

The last issue is the choice of the photometric profiles and the scale radius. We argue that the intermediate axis
is more appropriate for dynamics given the need to convert from an elliptical galaxy to
a spherical model.  The intermediate axis is smaller by $q^{1/2}$ than the major axis, where $q$
is the axis ratio, making the stellar distribution more compact which will generically produce higher central dispersions for the same mass. This is now the standard choice for \cite{TDCOSMO2025}, but not in some earlier studies. Using the intermediate axis will drive
the mass models to be less centrally concentrated and will lower $H_0$ compared to models using the major axis. The 3D light distributions in \cite{TDCOSMO2025} are based on multicomponent fits to the 2D photometric profiles. This is a significant improvement with respect to earlier studies, but there are still testable questions. For example, four of the eight lens galaxies are modeled with double S\'ersic profiles with indices fixed to $n_1=4$ and $n_2=1$, while fits to nearby galaxies show significant spreads in their indices \citep[e.g.,][]{Lange2016}. In S\'ersic models, overestimating the S\'ersic index leads to a more centrally concentrated distribution and a higher value of $H_0$ and vice versa.

The existence of color gradients in early-type galaxies also leads to a more fundamental problem of determining which photometric profile is ``correct''. Based on a simple thought experiment, the correct profile is the one which weights the contributions of the different stellar populations leading to the color gradients by their equivalent width contributions to the spectra. This is not the same as using the light profile closest to the wavelength region of the absorption lines used to determine the velocity dispersion. This is an interesting challenge at the required levels of precision, since in the presence of multiple stellar populations with differing kinematics, the quantity $n$ (Eqn.~\ref{eqn:weighting}) appearing in the Jeans equation corresponds to no easily measured quantity like a photometric brightness profile. A first approach to assess the potential biases that this effect may cause could be to measure the velocity dispersions and light profiles at different wavelength ranges.

Finally, we discuss the statistical consequences of a lens population dominated by early-type galaxies.  Early-type galaxies are fairly homogeneous in their dynamical properties,
their photometric properties and their stellar populations.  There are
trends with both velocity dispersion and redshift, but early-type galaxies
of comparable redshift and velocity dispersion are likely more similar than
dissimilar.  While the homogeneity of early-type galaxies clearly is not perfect, assuming that
they are completely inhomogeneous will underestimate the error budget when combining $H_0$ inferences.  For example, many time-delay cosmography studies used an O-M anisotropy model assuming that the anisotropy radius $r_a$ is a 
random variable for each lens \citep[e.g.,][]{Wong2020} means that any contribution from the
uncertainties in the anisotropy model to the final uncertainties diminish 
with the number of systems as $N^{-1/2}$.  If the lens galaxies are 
dynamically homogeneous, then essentially one value of $r_a/s$ (i.e., anisotropy radius in
units of the scale radius), characterizes the entire population and the
contribution of the uncertainties in the anisotropy model to the final 
uncertainties are independent of the number of systems. While the O-M model is not used in the most recent studies, the basic concept of this source of systematic error holds for any choice of anisotropy model. Homogeneous
photometric profiles mean that the same argument holds for choices of
the photometric profile.  For example, \cite{Suyu2010} and \cite{Wong2017} note that
using a \cite{Jaffe1983} model instead of a \cite{Hernquist1990} model
changes $H_0$ results by only 1\%.  But for a homogeneous population that shift has a systematic sense and so shifts all the lenses collectively in the same 
direction not as a random shift from galaxy to galaxy. For the photometric
problem, this issue can be avoided by just using the measured photometric
profiles as in \cite{TDCOSMO2025}, to the extent that there are no systematic problems in using the photometric
models (which there are, see the discussion of color gradient above). 

The homogeneity of the stellar populations means
that two velocity dispersion template stars likely produce similarly
different estimates of velocity dispersions for different lenses (this
hypothesis can be tested relatively easily). This
means that the scatter produced by choices of template stars should not
be assigned as an additional uncertainty to the individual dispersion
measurements.  Instead, global results should be computed for each choice
of template star and then the final distribution should be marginalized over
the template stars \citep[e.g.,][]{Knabel2025}.  It is possible to build models that lie at neither
extreme, assuming a general correlation combined with a scatter in the
correlation and then marginalizing over the parameters of the correlation
model.  This procedure is commonly used in cluster cosmology to, for
example, model the relationship between mass and richness (e.g., \citealt{Costanzi2019}).

In this work, we have not considered the role of ellipticity and rotation at all beyond
suggesting that spherical dynamical models of elliptical galaxies should probably
use the intermediate axis scale length rather than the major axis (or, e.g., marginalize
over the minor to major scale length range).  The effects of asphericity are seen both 
in the elliptical shapes of the galaxies
and the presence of odd velocity moments ($\langle v\rangle$ and $h_3$, see,
e.g., \citealt{Emsellem2004},
\citealt{Arnold2014},
\citealt{Veale2017}). For
a density ellipticity of $\epsilon=1-q$, the squared velocity dispersion will generally
show changes on the scale of the ellipticity of the gravitational potential,
$\Delta \sigma^2 \sim \epsilon/3$. \cite{Huang2026} addressed the systematic effects on $H_0$ due to the projection of triaxial early-type galaxies and selection effects of lens galaxies and found biases on $H_0$ up to 2-4\% when a spherical Jeans Anisotropic Model is assumed for central velocity dispersion measurements.

Achieving a 2\% estimate of $H_0$ using time-delay gravitational lensing implies knowing the measured velocity dispersions with an absolute accuracy of 1\%, provided the mass models allow enough radial and angular flexibility (\citealt{Kochanek2020,Kochanek2021}). In this work, we studied in detail several potential sources of systematics in the dynamics of the lens galaxies employed for time-delay cosmography. As seen in the summary Table \ref{tab:summary}, most of these systematic errors can bias $H_0$ inferences by more than 2\%, although the sense of the total corrections may shift $H_0$ to higher or lower values. It will be challenging to control these sources of uncertainty at the required precision.

\section*{Acknowledgments}

We thank K. Gebhardt, R. van der Marel, and S. Tremaine for answering a broad range of questions.
CSK is supported by NSF grants AST-2307385 and 2407206. This research was supported by the grant PID2024-160091NB-C32 funded by MCIU/AEI/10.13039/501100011033 / FEDER, EU. JAM is also supported by the project of excellence PROMETEO CIPROM/2023/21 of the Conselleria de Educaci\'on, Universidades y Empleo (Generalitat Valenciana).

\appendix
\section{A Simple Dynamical Model}\label{app:model}

For illustrative purposes, we constructed a simple isotropic dynamical model
that shifts between a flat rotation curve and a constant mass-to-light ratio.
We give the stars a \cite{Hernquist1990}\footnote{In the process of checking
various results, we discovered that Eqn.~B7 of \cite{Hernquist1990} is missing
a multiplicative factor of $2\pi$.}  distribution as defined in
Eqn.~\ref{eqn:hernrho}
while an SIS has
\begin{equation}
   \rho_S = \frac{ \sigma^2 }{ 2 \pi G r^2 }.
\end{equation}

The Hernquist model mimics
a de Vaucouleurs model of effective radius $R_e$
reasonably well for $s=0.55 R_e$.
The isotropic Jeans equation can be solved analytically for both a constant mass-to-light ratio case and for a Hernquist stellar profile embedded in an SIS potential (flat rotation curve case),
with
\begin{equation}
\begin{split}
\sigma_H^2(r)&=\frac{ G M }{ 12 s} \bigg[
   12 \rh (1+\rh)^3 \ln(1+1/\rh)+\\
   &- \frac{ \rh }{ 1+\rh}
    (25 + 52 \rh + 42\rh^2 + 12 \rh^3 )\bigg] \equiv \frac{ G M }{ s } \hat{\sigma}_H^2(\rh)
\end{split}
   \label{eqn:herndisp}
\end{equation}and
\begin{equation}
\begin{split}
\sigma_S^2(r) &= \sigma^2 (1+\rh)\left[ 2 + 9 \rh + 6\rh^2
  - 6 \rh (1+\rh)^2\ln(1+1/\rh)\right]\\
  &\equiv\sigma^2\hat{\sigma}_S^2(\rh),
\end{split}
   \label{eqn:sisdisp}
\end{equation}respectively and $\rh$ defined as $\rh=r/s$.  The masses enclosed  by cylinders
of radius $R$ are
\begin{equation}
 M_H(<R) = M \Rh^2 \frac{ X(\Rh)-1 }{ 1-\Rh^2 } = M m(<\Rh)
\end{equation}where $\Rh=R/s$ and $X(x)=\hbox{sech}^{-1}(x)/(1-x^2)^{1/2}$
for $x < 1$ and $\hbox{sec}^{-1}(x)/(x^2-1)^{1/2}$ for 
$x > 1$, and
\begin{equation}
  M_S(<R) = \pi \sigma^2 R/G,
\end{equation}respectively. And the surface density for the Hernquist and SIS profiles, respectively, are
\begin{equation}
 \Sigma_H(R) = \frac{M}{2\pi s^2(1-\Rh^2)^2} \left[ (2+\Rh^2)X(\Rh)-3 \right] \equiv \frac{M}{s^2}\hat{\Sigma}_H(\Rh)
\end{equation}and
\begin{equation}
  \Sigma_S(R) = \sigma^2/(2GR).
\end{equation}

We normalize the models by the projected mass inside the Einstein ring,
$M_E$. For a pure SIS model, $M_E=\pi \sigma_{SIS}^2 R_E/G$ and for a
pure Hernquist model $M_E= M_0 m(<R_E)$. We parametrize the
models by the fraction of the mass $f$ inside the Einstein ring from the
SIS model, which means that $f=\sigma^2/\sigma_{SIS}^2$ and that $1-f =
M/M_0$ where $\sigma$ and $M$ are the current normalizations of the two
models.  As we vary $f$ from $f=1$ to $0$, $\sigma$ changes from
$\sigma_{SIS}$ to zero and $M$ changes from $0$ to $M_0$.  That the mass
inside the Einstein ring is the same for both $f=0$ and $f=1$ provides
the transformation between $\sigma_{SIS}$ and $M_0$.
Then the 3D velocity dispersion profile of the stars in a combination of both mass distributions is
\begin{equation}
  \frac{\sigma^2}{\sigma_{SIS}^2 } =
   f \hat{\sigma}_S^2(\rh) + (1-f) \frac{ \pi \Rh_E }{ m(<\hat{R}_E) } \hat{\sigma}_H^2(\rh) 
\end{equation}where $\hat{\sigma}_S^2(r)$ and 
$\hat{\sigma}_H^2(r)$ are the normalized velocity dispersions defined in Eqn. \ref{eqn:sisdisp} and \ref{eqn:herndisp}, respectively.  We can then characterize
the effects on the Hubble constant using the surface
density at the Einstein radius compared to the
$\kappa_E=1/2$ expected for an SIS,
\begin{equation}
  \frac{\kappa }{\kappa_{SIS} }
  = f + \frac{ 2 \pi (1-f) \hat{R}_E^2 }{ m(<\hat{R}_E) } \hat{\Sigma}_H(\hat{R}_E).
\end{equation}

\begin{figure}
\centering
\includegraphics[width=0.47\textwidth]{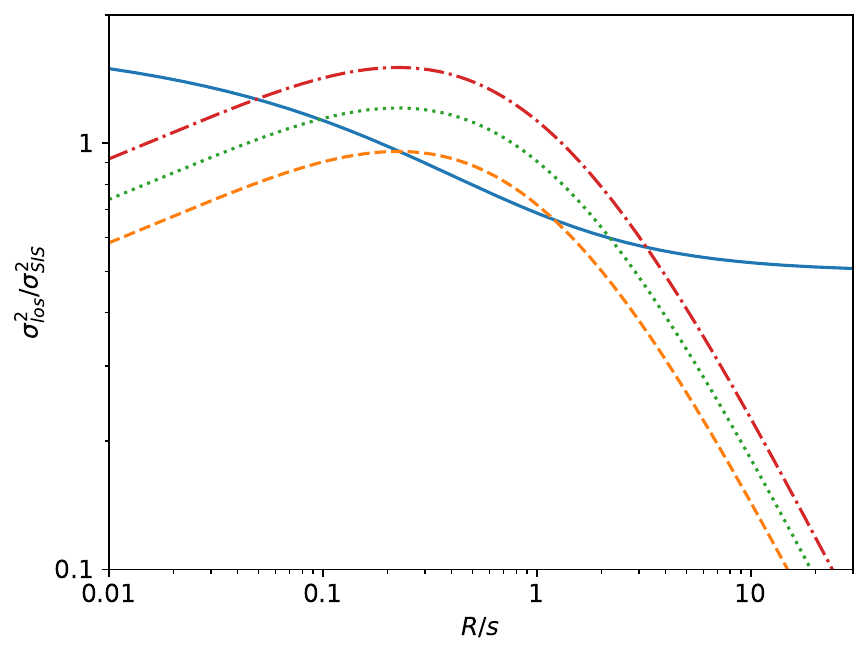}
\caption{
  The line-of-sight velocity dispersion profiles $\sigma_{los}^2$ of isotropic Hernquist models and a Hernquist model in the potential of an SIS (solid).  The Hernquist models are
  normalized to have the same projected mass as the SIS model inside $3s$ (top red dashdotted line), $2s$ (middle green dotted line)
  and $s$ (bottom orange dashed line).  The curves in Figure~\ref{fig:model} correspond to picking the right
  normalization, extracting the observed dispersion weighted by a Gaussian PSF model and the aperture, and then
  adding them weighted by $1-f_{SIS}$ and $f_{SIS}$, respectively.
  }
\label{fig:vdisp}
\end{figure}

Figure~\ref{fig:vdisp} shows the line-of-sight velocity dispersion
profiles for the two limiting isotropic cases.  They are normalized by the 
velocity dispersion corresponding to the SIS model and to have the
same enclosed mass at $R<s$, $R<2s$ or $R<3s$.  As the normalization
radius increases, the Hernquist mass distribution has more mass for
the same density profile, so its peak velocity dispersion increases
relative to the SIS model.  While these distributions include no
model for the effects of seeing, note how a model normalized to 
have the same $M(R<s)$ could easily show little variation in the
velocity dispersion with the mass fraction assigned to each model
if the FWHM of the PSF and the aperture favored spatial scales a similar to
$s$.  This is essentially what is occurring for RX~J1131$-$1231 and WGD~2038$-$4008 in
Figure~\ref{fig:model}. Whereas if the FWHM of the PSF and the aperture are small compared to $s$, the trend of $\sigma^2/\sigma^2_{SIS}$ with $f_{SIS}$ could even be negative, as HE~0435$-$1223 and WFI~2033$-$4723 exemplify.

To compute the los velocity distributions, we can simply use the
analytic Osipkov-Merritt distribution functions from \cite{Hernquist1990}
when there is no dark matter ($f_{SIS}=0$).  For the flat rotation
curve model ($f_{SIS}=1$), we have to compute the distribution functions
numerically, although we used a \cite{Jaffe1983} model with 
$\rho \propto r^{-2} (r+a)^{-2}$ to truncate the mass distribution
at large radii so that the escape velocities are finite at all
radii.  We set $a = 10^4 s$, so for the dynamics of the stars
it is effectively an SIS model.  For our illustrative purposes it
seemed unnecessary to model the los velocity distributions of the
intermediate cases, although the same numerical methods would 
work.  For the Abel integral needed to determine $f(Q)$ given
a complex potential,  it is important to remember that 
$d\rho_Q/d\Psi = (d\rho_Q/dr)/(d\Psi/dr)$!

\begin{table*}
  \centering
  \movetableright=-1in
  \caption{Photometric S\'ersic models for the lenses used in previous cosmological studies.}
  \begin{tabular}{lccccccccc}
  \hline
  \hline
  \multicolumn{1}{c}{Lens}    &
  \multicolumn{1}{c}{$n_1$} &
  \multicolumn{1}{c}{$A_1$} &
  \multicolumn{1}{c}{$R_{e1}$} &
  \multicolumn{1}{c}{$q_1$}   &
  \multicolumn{1}{c}{$n_2$} &
  \multicolumn{1}{c}{$A_2$} &
  \multicolumn{1}{c}{$R_{e2}$} &
  \multicolumn{1}{c}{$q_2$}   &
  \multicolumn{1}{c}{$R_{e}$} \\
  \multicolumn{1}{c}{}    &
  \multicolumn{1}{c}{} &
  \multicolumn{1}{c}{} &
  \multicolumn{1}{c}{(\arcsec)} &
  \multicolumn{1}{c}{}   &
  \multicolumn{1}{c}{} &
  \multicolumn{1}{c}{} &
  \multicolumn{1}{c}{(\arcsec)} &
  \multicolumn{1}{c}{}   &
  \multicolumn{1}{c}{(\arcsec)} \\
\hline
      DES~J0408$-$5354 & & & & &         &         &         &         &$ 1.200$ \\
    HE~0435$-$1223 &$ 0.813$ &$ 0.037$ &$ 1.953$ &$ 0.875$ &$ 1.887$ &$ 0.368$ &$ 0.312$ &$ 0.826$ &$ 1.330$ \\
     PG~1115$+$080 &$ 3.402$ &$ 0.752$ &$ 0.263$ &$ 0.870$ &$ 0.482$ &$ 0.049$ &$ 1.249$ &$ 0.984$ &$ 0.529$ \\
   RX~J1131$-$1231 &$ 0.930$ &$ 0.091$ &$ 2.490$ &$ 0.878$ &$ 1.590$ &$ 0.890$ &$ 0.362$ &$ 0.849$ &$ 1.850$ \\
 SDSS~J1206$+$4332 &$ 2.660$ &$ 1.000$ &$ 0.600$ &$ 0.890$ &         &         &         &         &$ 0.600$ \\
      B1608$+$656 &$ 4.000$ &$ 1.000$ &$ 0.580$ &$ 1.000$ &         &         &         &         &$ 0.580$ \\
   WFI~2033$-$4723 &$ 1.454$ &$ 1.724$ &$ 0.252$ &$ 0.756$ &$ 1.194$ &$ 0.190$ &$ 1.893$ &$ 0.787$ &$ 1.410$ \\
    WGD~2038$-$4008  & & & & &         &         &         &         &$ 2.400$ \\
\hline
\end{tabular}
\tablecomments{The parameters $n$, $A$, $R$, and $q$ correspond to those of Eqn. \ref{eqn:sersic} and the subindices reference the first and second (if applied) profiles that fit the galaxy lens light distribution. The total effective radius is listed in the last column. The light model parameters, when shown, came from a private communication if not explicitly reported in the corresponding modeling paper.}
  \label{apptab:phot}
\end{table*}

\begin{table*}
  \centering
  \movetableright=-1in
  \caption{Summary of dispersion observations for the time-delay lenses.}
  \begin{tabular}{lDcDr@{$\times$}lcc}
  \hline
  \hline
  \multicolumn{1}{c}{Lens}    &
  \multicolumn{2}{c}{$z_{\text{lens}}$}   &
  \multicolumn{1}{c}{$\sigma_{\ast}$}   &
  \multicolumn{2}{c}{Seeing} &
  \multicolumn{2}{c}{Aperture}   &
  \multicolumn{1}{c}{s}  &
  \multicolumn{1}{c}{$\theta_E$} \\
  \multicolumn{1}{c}{}    &
  \multicolumn{2}{c}{}   &
  \multicolumn{1}{c}{(km/s)}   &
  \multicolumn{2}{c}{FWHM (\arcsec)} &
  \multicolumn{2}{c}{(\arcsec$^2$)}   &
  \multicolumn{1}{c}{(\arcsec)}   &
  \multicolumn{1}{c}{(\arcsec)}   \\
\hline
\decimals
DES~J0408$-$5354 & 0.60 & 220 & 0.52 & 0.75&1.00 & 0.66 & 1.92 \\
HE~0435$-$1223     &0.45  &222 &0.8 & 0.75&0.54   &0.73 &1.18 \\
PG~1115$+$080     &0.31  &281 &0.8 &1.00&1.06   &0.29 &1.04 \\
RX~J1131$-$1231    &0.30  &323 &0.7 &0.70&0.81   &1.02 &1.64 \\
SDSS~J1206$+$4332  &0.75  &290 &1.0 &1.00&1.90   &0.33 &1.57\\
B1608$+$656       &0.63  &260 &0.9 &1.00&0.84   &0.32 &0.81  \\
WFI~2033$-$4723    &0.66  &250 &1.0 &1.80&1.80   &0.78 &0.93 \\
WGD~2038$-$4008 & 0.23 & 296 & 0.90 & 0.75&1.00 &1.32 &1.38 \\
\hline
\end{tabular}
\\
\tablecomments{We list the lens redshift, $z_{\text{lens}}$, the measured velocity dispersion, $\sigma_*$, the seeing reported as the full width at half maximum (FWHM), the extraction aperture, the Hernquist scale radius, $s=0.55R_e$, and the Einstein radius, $\theta_E$.}
  \label{apptab:data}
\end{table*}

\section{Previous velocity dispersion measurements}\label{app:prev}

Prior to the observations reported in \cite{TDCOSMO2025}, ground-based spectroscopy of the eight lenses were used to infer $H_0$ in \cite{Wong2020}, \cite{Birrer2020}, and \cite{Wong2024}. Although the new (mostly space-based) IFU measurements are going to be used for future cosmological inferences, it is still interesting to review the level of potential systematics errors for these previous works and compare it with the current one. In Table~\ref{apptab:phot} we report the properties of the \cite{Sersic1968}
models as these were generally preferred over the pseudo-Jaffe models.

For the systems where double \cite{Sersic1968} models are available, the effective 
radius is estimated by computing the fraction of the light inside circular apertures
and setting the effective radius to encompass half of the projected light.
For the single \cite{Sersic1968} models, the reported major axis effective
radius appears to have been used when modelling the dynamical constraints.
For DES~J0408$-$5354 and WGD~2038$-$4008 only the value of $R_e$ is reported by \cite{Birrer2020} and \cite{Shajib2023}.

The velocity dispersion for DES~J0408$-$5354 is from \cite{BuckleyGeer2020}. There are eight different measurements based on fits using stellar population models from MILES stellar templates (\citealt{SanchezBlazquez2006,FalconBarroso2011}). Four of these were averaged and used in the lens models (\citealt{Shajib2020,Birrer2020}). For reference, we study the systematics of the measurement with the smallest aperture and the best seeing, namely the observations with Gemini South GMOS-S R400 grating. The posterior central values from \cite{Birrer2020} are adopted for $R_e$ and $\theta_E$.

\begin{figure*}
\centering
\includegraphics[width=\textwidth]{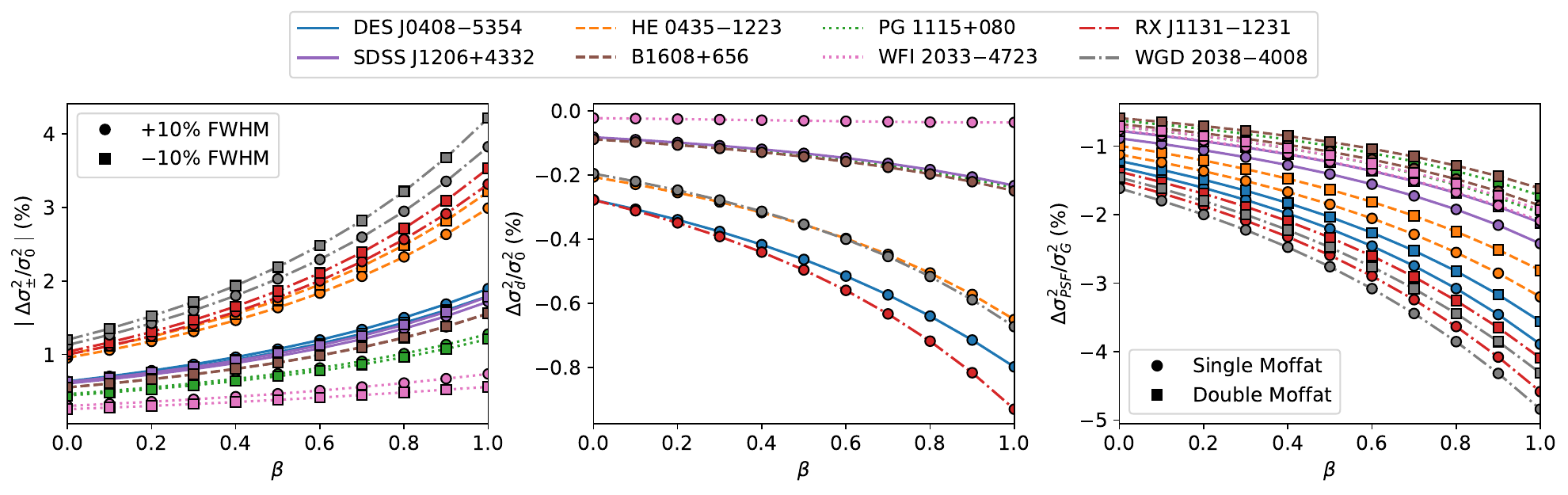}
\caption{
  Fractional changes in  $\sigma^2$, as a function of anisotropy, $\beta$, due to changes in the seeing FWHM (left), miscentering the aperture by 0\farcs1 (center), and changes in the PSF profile (right).
  }
\label{appfig:psf}
\end{figure*}

\begin{figure*}
\centering
\includegraphics[width=\textwidth]{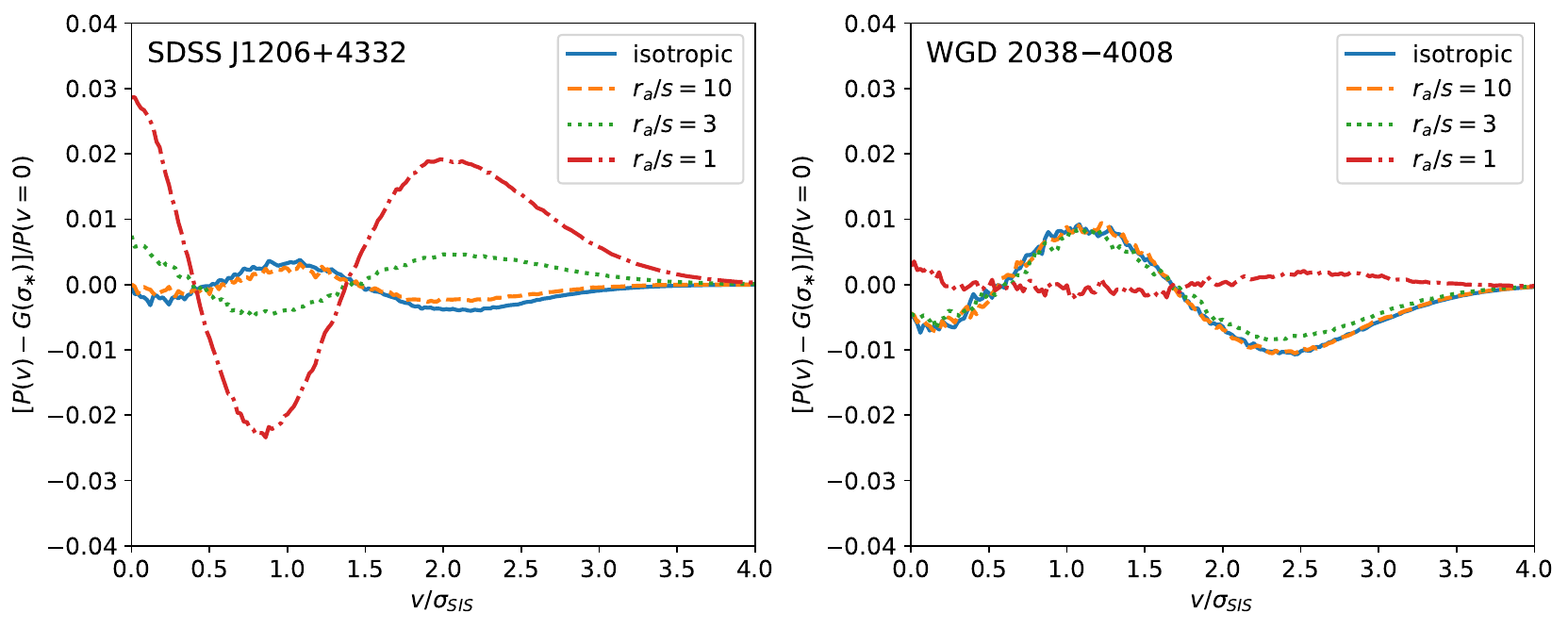}
\caption{
  The differences $[P(v)-G(\sigma_{\ast})]/P(v=0)$ 
  between the los velocity distribution ($P(v)$) and the best 
  fit Gaussian model ($G(\sigma_{\ast})$) normalized by the peak of
  the los velocity distribution at zero velocity, $P(v=0)$, 
  for SDSS~J1206$+$4332 (left) and WGD~2038$-$4008 (right). The anisotropy radii considered are 
  $r_a/s\rightarrow \infty$ (isotropic), $10$, $3$, and $1$. 
  The velocity distribution accounts for Gaussian PSF effects and the rectangular extraction aperture from Table~\ref{apptab:data} is applied. There is some noise 
  due to the Monte Carlo integration and sampling.
  }
\label{appfig:DFs}
\end{figure*}

\begin{figure*}
\centering
\includegraphics[width=\textwidth]{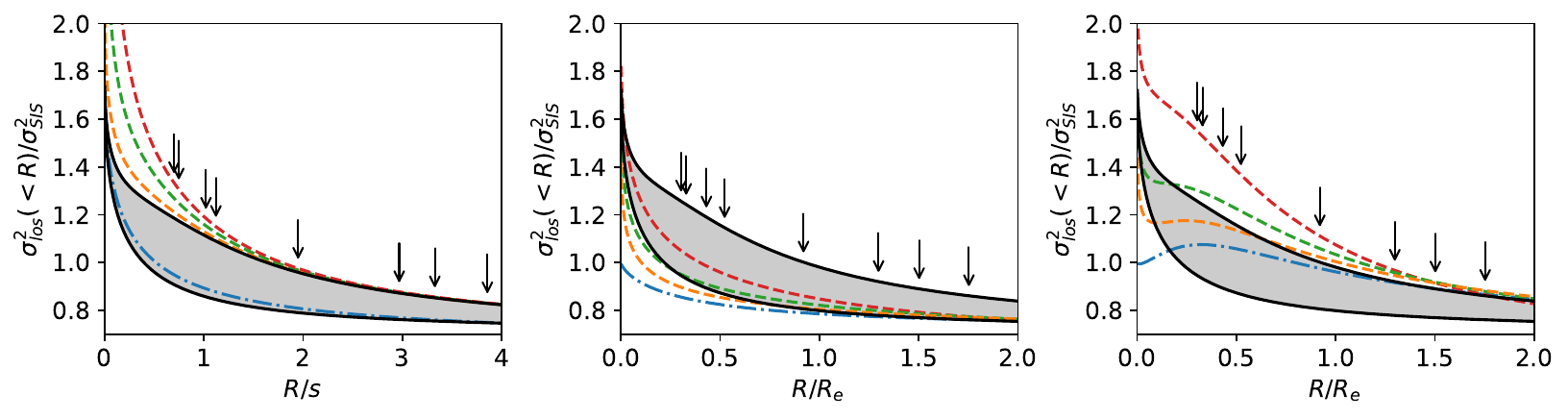}
\caption{\textbf{Left:} Mean squared los velocity dispersion inside radius $R$, $\sigma_{los}^2(<R)$, for the Hernquist profile embedded in an SIS mass model normalized by the squared velocity dispersion $\sigma_{SIS}^2$. The solid lines are for the O-M model with $r_a/s=10$ (bottom) and $r_a/s=1$ (top) and the shaded region is the allowed range of these models. The blue dashdotted line is for the anisotropy profile given in Eqn.~\ref{eqn:df2} with $r_a/s=2.5$. The dashed lines are for the anisotropy profile given in Eqn.~\ref{eqn:df3} with $r_a/s=1$ and $\beta_0=0.1$ (lower orange), $0.3$ (middle green) and $0.5$ (upper red). \textbf{Center and right:} Mean squared enclosed los velocity dispersion, 
$\sigma_{los}^2(<R)/\sigma_{SIS}^2$, for various stellar
density distributions normalized to the same effective radius $R_e$ embedded in an SIS mass distribution and using the O-M anisotropy model. The solid lines and the shaded region are the range that
the \protect\cite{Hernquist1990} profile can produce and the blue dashdotted 
lines are for the \protect\cite{Jaffe1983}  profile. The 
dashed lines are for the \protect\cite{Sersic1968} 
profiles with $n=4$ 
(orange), $n=3$ (green) and $n=2$ (red). The models on the center (right) panel have an anisotropy radius of $r_a=0.5R_e$ ($r_a=5R_e$) except for the \protect\cite{Hernquist1990} profile where both anisotropy profiles are shown for reference. The arrows are are placed at the effective circular aperture radius of the eight lenses in the order from left to right of WGD~2038$-$4008, RX~J1131$-$1231, HE~0435$-$1223, DES~J0408$-$5354, WFI~2033$-$4723, B1608$+$656, PG~1115$+$080, and SDSS~J1206$+$4332 for the three panels.
  }
\label{appfig:aniphot}
\end{figure*}

The velocity dispersion for HE~0435$-$1223 comes from \cite{Wong2017}.
The reported uncertainty of $15$~km/s is the quadrature sum of a
$11$~km/s statistical contribution and a $10$~km/s contribution from
the scatter between results for different templates.  The 
templates used in the H0LiCOW analyses are drawn from the INDO-US
templates (\citealt{Valdes2004}).  The standard set of stars are
the roughly Solar metallicity giants
HD102328 (K3III), HD107950 (G5III), HD111812 (G0III), HD115604 (F2III), 
HD124897 (K1III), HD148387 (G8III), HD163588 (K2III), HD168723 (K0III), 
and HD188350 (A0III), with the fits done including and excluding the
earlier A0 and F2 stars (Auger, private communication). Sky lines
are used to correct for the estimated instrumental resolution of 37~km/s.  
The HST-based photometric model from \cite{Wong2017} was used
instead of the AO-based models in \cite{Chen2019}.
The Einstein radius is estimated to be $\theta_E=1\farcs18$.

The velocity dispersion for PG~1115$+$080 comes from \cite{Tonry1998}. The extracted region is reported as 1\farcs0, which probably
means 5 Keck/LRIS pixels or 1\farcs06.  Two stars (HD~132737 and AGK2 $+14.873$)
were used as radial velocity templates. \cite{Chen2019} simply used
the \cite{Tonry1998} dynamical results, so this velocity dispersion does not
include the range of stellar template uncertainties of the other
systems.  They also seem to have assumed a $1\farcs0 \times 1\farcs0$ 
aperture rather than the $1\farcs0 \times 1\farcs06$ resulting from 
extracting 5 pixels. We used the photometric model from \cite{Chen2019}
and their estimated Einstein radius of $\theta_E=1\farcs18$.

The velocity dispersion for RX~J1131$-$1231 comes from \cite{Suyu2013}.
The analysis procedures are the same as for HE~0435$-$1223.  The
photometric model is from \cite{Suyu2013}  and they report an
Einstein radius of $\theta_E=1\farcs64$.  

The velocity dispersion for SDSS~J1206$+$4332 comes from \cite{Agnello2016} and the seeing FWHM is from Agnello (private communication). The lens photometry is modeled with a single
\cite{Sersic1968} model.
They find an Einstein radius of $\theta_E=1\farcs57$ for their model.

The velocity dispersion for B1608$+$656 comes from \cite{Suyu2010}. The
\cite{deVaucouleurs1948} effective radius is estimated to be $R_e=0\farcs58$ ($s=0\farcs32$) from
\cite{Koopmans2003}. This is a circularized effective radius.
We adopt the Einstein radius estimate of $\theta_E=0\farcs81$ from
\cite{Suyu2009}.

The velocity dispersion for WFI~2033$-$4723 is from \cite{Rusu2020}. These were integral field unit (IFU) spectra and so 
are less affected by some of the PSF issues such as measuring the seeing accurately. 
The photometric model is also from \cite{Rusu2020}, and they use
an Einstein radius of $\theta_E=0\farcs93$. 

The velocity dispersion for WGD~2038$-$4008 is also from \cite{BuckleyGeer2020} using MILES stellar templates. Here we use the measurement later used in the mass model (\citealt{Shajib2023}). It was obtained with the GMOS-S B600 grating at Gemini South. The lens was modeled with two different packages, \textsc{glee} and \textsc{lenstronomy} (\citealt{Shajib2023}). We use the \textsc{lenstronomy} inferences since they explicitly report both $R_e$ and $\theta_E$.

\begin{table*}
    \caption{Summary of the maximum percentage changes in the squared velocity dispersion, $\Delta\sigma^2/\sigma^2$, due to possible sources of systematic errors for the eight time-delay lenses under the conditions of previous measurements and mass models from Table~\ref{apptab:data}.}
    \label{apptab:summary}
    \centering
    \movetableright=-1in
    \resizebox{\textwidth}{!}{%
    \begin{tabular}{lSSSSSSSS}
    \hline
    \hline
    $\Delta\sigma^2/\sigma^2$ (\%) & {DES~J0408} & {HE~0435} & {PG~1115} & {RX~J1131} &  {SDSS~J1206} & {B1608} &  {WFI~2033} & {WGD~2038}\\
    \hline
    \textsc{Point Spread Function} & & & & & & & & \\
 Larger seeing & {$-$}1.9& {$-$}3.0& {$-$}1.3& {$-$}3.3& {$-$}1.7& {$-$}1.6& {$-$}0.7& {$-$}3.8\\
 Smaller seeing & {$+$}1.8& {$+$}3.2& {$+$}1.2& {$+$}3.5& {$+$}1.8& {$+$}1.6& {$+$}0.6& {$+$}4.2\\
 Miscentering & {$-$}0.8& {$-$}0.7& {$-$}0.2& {$-$}0.9& {$-$}0.2& {$-$}0.2& {$-$}\hspace{-0.5em}0.04& {$-$}0.7\\
 PSF Single Moffat & {$-$}3.9& {$-$}3.2& {$-$}2.0& {$-$}4.6& {$-$}2.4& {$-$}1.9& {$-$}2.1& {$-$}4.8\\
 PSF Double Moffat & {$-$}3.6& {$-$}2.8& {$-$}1.7& {$-$}4.1& {$-$}2.1& {$-$}1.6& {$-$}1.9& {$-$}4.3\\
 \hline
    \textsc{Measured Dispersion} & & & & & & & & \\
 Velocity DF & {$+$}3.8& {$+$}4.1& {$-$}8.4& {$+$}4.6& {$-$}10.5& {$-$}6.9& {$-$}4.3& {$+$}4.7\\
 \hline
    \textsc{Anisotropy model} & & & & & & & & \\
 Cuddeford $\beta_0$=0.1 & {$+$}1.1& {$+$}1.2& {$+$}0.1& {$+$}2.0& {$+$}0.1& {$+$}0.1& {$+$}0.3& {$+$}2.2\\
 Cuddeford $\beta_0$=0.3 & {$+$}3.3& {$+$}3.9& {$+$}0.3& {$+$}6.4& {$+$}0.2& {$+$}0.4& {$+$}1.0& {$+$}7.0\\
 Cuddeford $\beta_0$=0.5 & {$+$}5.6& {$+$}6.7& {$+$}0.4& {$+$}11.3& {$+$}0.3& {$+$}0.6& {$+$}1.7& {$+$}12.5\\
 \hline
    \textsc{Photometric model} & & & & & & & & \\
 Jaffe & {$-$}5.2& {$-$}6.6& {$-$}0.3& {$-$}8.8& {$+$}0.2& {$-$}0.7& {$-$}2.0& {$-$}9.5\\
 S\'ersic $n$=4 & {$-$}1.9& {$-$}3.0& {$+$}2.7& {$-$}4.7& {$+$}2.5& {$+$}2.8& {$+$}2.1& {$-$}5.2\\
 S\'ersic $n$=3 & {$+$}6.1& {$+$}5.5& {$+$}3.1& {$+$}4.1& {$+$}2.1& {$+$}4.0& {$+$}5.7& {$+$}3.6\\
 S\'ersic $n$=2 & {$+$}19.3& {$+$}21.2& {$+$}2.7& {$+$}23.0& {$+$}0.5& {$+$}5.1& {$+$}10.9& {$+$}23.3\\
    \hline
    \end{tabular}}
\end{table*}

We show in Figure~\ref{appfig:psf} the fractional changes in $\sigma^2$ produced for the PSF related systematic errors considered in Sect.~\ref{sec:PSF}. The model assumptions and direction of the corrections are the same as in Sect.~\ref{sec:PSF}. RX~J1131$-$1231 and WGD~2038$-$4008 nearly always have the largest changes 
because they are large lenses ($s=1\farcs02$ and $s=1\farcs32$, respectively) observed in
small apertures ($0\farcs70 \times 0\farcs81$ and $0\farcs75 \times 1\farcs00$) so small changes in the aperture or PSF have a larger impact on the measured velocity dispersion.

For the differences in the measured velocity dispersion with respect to the true $v_\text{rms}$ (Figure~\ref{appfig:DFs}), SDSS~J1206$+$4332 still defines the largest negative differences whereas WGD~2038$-$4008 has the strongest positive differences because this was the largest lens compared to its aperture for the previous set of observations and models (Table~\ref{apptab:data}).

The left panel in Figure~\ref{appfig:aniphot} shows the differences of different anisotropy models with respect to the O-M model with with $r_a/s=10$ and $r_a/s=1$, since this was the preferred anisotropy model in previous time-delay cosmological inferences. As expected, the largest differences are produced by the \cite{Cuddeford1991} model with $\beta_0=0.5$ as in Sect.~\ref{sec:isotropy}.

The differences in the choice of the photometric model (center and right panel in Figure~\ref{appfig:aniphot}) are also computed under the assumption of O-M anisotropic model with the same range of anisotropy radii. The S\'ersic $n=2$ profile spans larger stellar dispersions outside the envelope of the Hernquist model as in Sect.~\ref{sec:profile}. Photometric models that produce velocity dispersions inside those spanned by the Hernquist model will not induce large differences in the dynamical model. A clear practical example can be found in \cite{Suyu2010} and \cite{Wong2017}. They used both a
Hernquist and a Jaffe model for the stellar distribution in
their models of B1608$+$656 and HE~0435$-$1223, respectively and
found that it made almost no difference.  From Figure~\ref{appfig:aniphot}
we can see that this is expected because the range of
velocity dispersions allowed by the models are almost 
identical for the aperture
sizes of the observations relative to $R_e$.

In Table~\ref{apptab:summary}, we summarize the maximum fractional changes for each lens for all the sources of systematic errors considered using the observational parameters in Table~\ref{apptab:data}. The directions of the corrections for each source of systematics are the same as reported for the new measurements. In comparison to Table~\ref{tab:summary}, the absolute changes in velocity dispersion produced by the PSF model and aperture uncertainties are larger for the previous set of observations. The fractional changes between the measured velocity dispersion and the actual $v_\text{rms}$ tends to be more positive for the new observations because the apertures are generally smaller compared to the scale radii of the lenses. For the same reason, the impact of the assumed anisotropy and photometric model is larger for the new set of velocity dispersion measurements (except for the extreme cases in these studies where S\'ersic profiles with $n=2$ or 3 were modeled with a Hernquist mass distribution).

\bibliography{ms2}{}
\bibliographystyle{aasjournalv7}

\end{document}